\documentclass[10pt,aps,pre,twocolumn,superscriptaddress,showpacs]{revtex4-1}
\usepackage{enumerate,amssymb,ascmac,amsmath,bm}
\usepackage[dvipdfmx]{graphicx}
\usepackage[dvipdfmx]{color}
\newcommand{\dis}{\displaystyle}         

\begin{document}

	\title{Eigenvalue analysis of an irreversible random walk \\
	with skew detailed balance conditions}
	
	\author{Yuji Sakai}
	\email[]{yuji0920@huku.c.u-tokyo.ac.jp}
	\affiliation{Graduate School of Arts and Sciences, 
	The University of Tokyo, 
	3-8-1 Komaba, Meguro-ku, Tokyo 153-8902, Japan}
	
	\author{Koji Hukushima}
	\email[]{hukusima@phys.c.u-tokyo.ac.jp}
	\affiliation{Graduate School of Arts and Sciences, 
	The University of Tokyo, 
	3-8-1 Komaba, Meguro-ku, Tokyo 153-8902, Japan}
	\affiliation{Center for Materials Research by Information Integration,
	National Institute for Materials Science, 1-2-1 Sengen, Tsukuba, 
	Ibaraki 305-0047, Japan}

	\date{\today}

	\begin{abstract}
	An irreversible Markov-chain Monte Carlo (MCMC) algorithm
	with skew detailed balance conditions 
	originally proposed by Turitsyn et al. 
	is extended to general discrete systems on the basis of the
	Metropolis-Hastings scheme.
	To evaluate the efficiency of our proposed method,
	the relaxation dynamics of 
	the slowest mode and the asymptotic variance are studied
	analytically in a random walk on one dimension. 
	It is found that the performance in irreversible MCMC
	methods violating the detailed balance condition is
	improved by appropriately choosing parameters in the algorithm. 
	\end{abstract}


	\maketitle


	\section{\label{Sect:Introduction}Introduction}
        
	Markov-chain Monte Carlo (MCMC) methods have already been 
	applied to numerous problems in various fields such as 
	physics, biochemistry, information sciences, 
	and economics~\cite{Landau, Liu}.
	Ever since Metropolis et al. 
	invented the MCMC method in 1953~\cite{Metropolis1953}, 
	many kinds of improved MCMC methods have been proposed and some of them 
	have contributed to the development of sciences, 
	especially to the understanding of phase transitions and 
	critical phenomena in statistical physics.
        
	Applying an MCMC method to a problem requires the preparation of
	a target distribution and a transition matrix in a Markov chain. 
	For ascertaining the efficiency of MCMC methods,
	it is important to investigate whether 
	the distribution in a Markov chain converges to the stationary
	target distribution 
	rapidly and whether the correlation between samples from 
	the Markov chain is sufficiently small. 
	The multicanonical method~\cite{Berg1992} and 
	the replica-exchange Monte Carlo 
	method~\cite{KH1996} improve 
	the convergence rate quantitatively 
	by extending the target distribution.
	Cluster algorithms, 
	such as the Swendsen-Wang algorithm~\cite{Swendsen1987} 
	and the Wolff algorithm~\cite{Wolff1989}, yield a remarkable reduction 
	in the critical slowing down 
	of spin systems by choosing an appropriate transition matrix 
	with a multi-spin update. 
	The correlation in a Markov chain, 
	which is worth considering in addition to 
	the convergence rate, is characterized by an asymptotic variance.
	This asymptotic variance can be reduced by decreasing the rejection 
	rate in a Markov chain in MCMC methods 
	with the detailed balance condition (DBC); 
	this is known as Peskun's theorem~\cite{Peskun1973}.
	As mentioned in the following sections, both the convergence rate and 
	the asymptotic variance are closely related to the second-largest 
	eigenvalue of the transition matrix. 
	Thus, improving the efficiency in MCMC 
	methods is equivalent to reducing the second-largest eigenvalue 
	of the transition matrix in the corresponding Markov chain.

	The detailed balance condition is usually imposed upon the transition 
	matrix to guarantee that our target distribution is exactly 
	the stationary distribution in the Markov chain. 
	Conventional MCMC methods, such as the Metropolis-Hastings 
	algorithm~\cite{Hastings1970} and 
	heat-bath method~\cite{Barker1965}, 
	have been developed within the framework of the DBC.
	However, it is possible to construct 
	such a Markov chain without imposing the 
	DBC because the DBC is not always necessary but sufficient to make the 
	MCMC method work correctly. Furthermore, it is unclear whether 
	MCMC methods with the DBC are more efficient 
	than those without the DBC.
	In fact, several studies have shown numerically and partly
	analytically that the efficiency of 
	MCMC methods can be improved by violating the DBC in some 
	cases~\cite{Ren2006, Suwa-Todo2010, Todo2013, 
	Diaconis1997,Chen1999, Turitsyn2011, 
	Fernandes2011, Schram, YS2013, KH2013, 
	Bernard2009, Michel2014, Michel2015, Nishikawa2015}. 
	Hence MCMC methods without the DBC have been eagerly studied recently.

	Strategies used to violate the DBC are roughly divided into two types.
	One is to reduce the rejection rate in the Markov chain, 
	as proposed by 
	Suwa and Todo~\cite{Suwa-Todo2010,Todo2013}. 
	They constructed a Markov chain without the DBC by 
	using a geometric allocation approach, showing numerically that 
	their method reduces the autocorrelation time by a factor of 
	more than 6 
	for four- and eight-state Potts models at the transition temperature.
	The other strategy is to extend the state space and the target 
	distribution~\cite{Diaconis1997,Chen1999, Turitsyn2011, 
	Fernandes2011, Schram, YS2013, KH2013, 
	Bernard2009, Michel2014, Michel2015, Nishikawa2015}.
	Such a strategy is called ``lifting,'' 
	and an irreversible Markov chain on the extended state space is 
	referred to as a ``lifted'' Markov chain.
	In Refs.~\cite{Diaconis1997} and \cite{Chen1999} it is shown that 
	the convergence toward the target distribution is accelerated by 
	applying the methodology of a ``lifted'' Markov chain 
	to a simple random walk. 
	Especially in Refs.~\cite{Turitsyn2011, Fernandes2011, Schram}, 
	the authors have reported that the dynamical critical exponent in 
	a two-dimensional Ising model and fully connected Ising model 
	can be reduced by their methods.
	The event-chain Monte Carlo (ECMC) algorithm~\cite{Bernard2009} 
	is constructed for systems of continuous degree of freedom, 
	such as hard-sphere, more general particle 
	systems~\cite{Michel2014}, and continuous 
	spin systems~\cite{Michel2015,Nishikawa2015}. 
	Recently, Nishikawa et al. have shown numerically that 
	the ECMC algorithm for a three-dimensional Heisenberg model reduces
	the dynamical critical exponent down to 
	$z\simeq 1$~\cite{Nishikawa2015}.
	These results imply that the violation of the DBC 
	could change the relaxation dynamics 
	of physical quantities qualitatively.

	The efficiency of the violation of the DBC is 
	partially confirmed theoretically. 
	Ichiki and Ohzeki have revealed that 
	the real part of the second-largest eigenvalue of a transition matrix 
	decreases by violating the DBC, compared to that of a symmetrized
	transition matrix satisfying the DBC~\cite{Ichiki2013}.
	The asymptotic variance is also reduced in comparison with that of the 
	corresponding symmetrized matrix~\cite{Sun2010}.
	It should be noted that the symmetrized transition matrix is not always
	equivalent to that before violating the DBC. Hence,
	the relation between the efficiency 
	of MCMC methods and the violation of the DBC in general 
	is not yet well understood.

	In this paper, we focus on the skew detailed balance condition (SDBC) 
	originally proposed by Turitsyn et al.~\cite{Turitsyn2011},
	which belongs to the latter strategy mentioned above. 
	We develop an irreversible Metropolis-Hastings algorithm with the SDBC 
	so that it can be applied to any general system.
	In general, it is quite difficult to evaluate the second-largest
	eigenvalue of the transition matrix of 
	a Markov chain even with the DBC.
	It is worth evaluating the efficiency of the algorithm in a toy
	model where all the eigenvalues of the matrix is explicitly
	written down under both the DBC and the SDBC.
	Here by applying 
	the proposed algorithm to a simple random walk on one dimension 
	the efficiency of the algorithm is studied. 
	Then, we show analytically that 
	the second-largest eigenvalue in absolute value and 
	the asymptotic variance of 
	the corresponding transition matrix 
	for the random-walk problem 
	can be reduced by imposing the SDBC.
	Our results imply that 
	the relaxation dynamics in a Markov chain can be qualitatively 
	changed from diffusive to ballistic by introducing the violation
	of the DBC,
	and that the violation does not always improve the efficiency. 
        
	The paper is organized as follows. 
	Section \ref{Sect:MCwithDBC} introduces the theoretical 
	foundation of a Markov chain with the DBC. 
	In Sec.~\ref{Sect:MCwithSDBC}, a Markov chain with the SDBC 
	is constructed.
	In Sec.~\ref{Sect:IMH}, an algorithm to realize 
	the Markov chain with the SDBC is proposed.
	In Sec.~\ref{Sect:Performance Evaluation}, 
	the efficiency of the proposed algorithm is discussed by analyzing 
	a random walk in one dimension. 
	Section \ref{Sect:Summary and Discussion} summarizes this study. 
        

	\section{\label{Sect:MCwithDBC}
	Markov Chain with the Detailed Balance Condition}

	The Markov chain with the SDBC is 
	constructed based on the Markov chain with the DBC. 
	We briefly review the theoretical aspects 
	of the Markov chain satisfying the DBC~\cite{Bremaud}.
	Note that all the vectors 
	in the paper are defined as row vectors.

	\subsection{\label{subsect:setup}Setup}
        
	Throughout the paper, we discuss a system of 
	discrete degree of freedom.
	Let $I=\{1,\ldots,\Omega\}$ be a state space of the system 
	with $\Omega$ being the total number of states.
	Suppose that a target probability distribution is given as 
	$\bm{\pi}=(\pi_1,\ldots,\pi_{\Omega})$ with $\pi_i>0$ and 
	$\sum_{i=1}^{\Omega}\pi_i=1$.

	We need to   
	(i) generate sampling states according to the target distribution and 
	(ii) calculate the expectation of a 
	quantity $\hat{f}$ with respect to the target distribution, 
	defined as 
		\begin{align}
		\langle\hat{f}\rangle_{\bm{\pi}}
		\equiv\sum_{i=1}^{\Omega}\pi_i f_i
		=\bm{\pi}\bm{f}^{\top},
		\end{align}
	where $f_i$ depending on the state $i$ denotes the realization of 
	$\hat{f}$ and $\bm{f}\equiv(f_1,\ldots,f_{\Omega})$.
	It is, however, difficult to evaluate 
	$\langle\hat{f}\rangle_{\bm{\pi}}$ 
	analytically for systems with  
	high-dimensional state space in general.
	Thus, MCMC methods are often employed 
	to achieve our aims numerically 
	for sufficiently large $\Omega$.
        

	\subsection{\label{subsect:Transition matrix and master equation}
	Transition matrix and master equation}
        
\begin{figure}[tb]
		\centering
		\includegraphics[width=.47\textwidth]{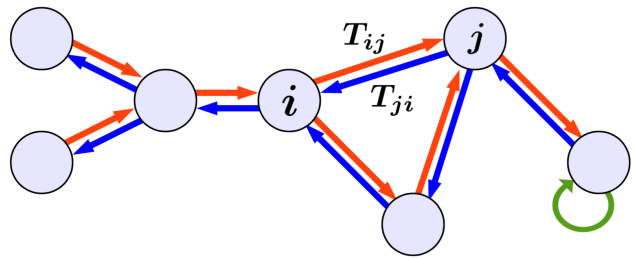}
		\caption{(Color online) Graphical representation of 
		the transition matrix.}
		\label{fig:statespace}
		\end{figure}

	A discrete-time Markov chain is a fundamental 
	stochastic process in which the transition probability from 
	the current state $i$ to a new state $j$ is independent of 
	the history of the transition.
	Let $\mathsf{T}=(T_{ij})_{i,j\in I}$ be the transition matrix of 
	the Markov chain. An element 
	$T_{ij}$ denotes the transition probability 
	from state $i$ to $j$ in a unit time of the Markov chain 
	(Fig.~\ref{fig:statespace}).
	Note that $\mathsf{T}$ is the stochastic matrix, {\rm i.e.}, 
	$T_{ij}\ge 0$ for all $i,j\in I$ and $\sum_{j=1}^{\Omega}T_{ij}=1$ 
	for all $i\in I$. 
        
	Let $\bm{p}^{(n)}\equiv(p_1^{(n)},\ldots,p_{\Omega}^{(n)})$ 
	be a probability distribution after $n$ steps in the Markov chain. 
	$p_i^{(n)}$ denotes the probability 
	for finding a state $i$ at time $n$. 
	Then, the time evolution of the probability distribution is 
	described by the master equation expressed as
		\begin{align}
		\bm{p}^{(n+1)}=\bm{p}^{(n)}\mathsf{T},
		\end{align}
	and, consequently, we obtain 
	\begin{align}
	\bm{p}^{(n)}=\bm{p}^{(0)}\mathsf{T}^n,
	\end{align}
	which shows that the distribution at arbitrary time is completely 
	characterized by an initial distribution and the transition matrix. 

                
	\subsection{\label{Detailed balance condition}
	Detailed balance condition}
        
	A transition matrix $\mathsf{T}$ is called ergodic 
	if there exists an integer $n>0$ 
	such that all elements of the $n$th power of 
	the transition matrix are positive. 
	A Markov chain characterized by an ergodic transition matrix is 
	ensured to have a unique stationary distribution.
	It is guaranteed that $\bm{p}^{(n)}$ converges to our desired 
	distribution $\bm{\pi}$ as $n\to\infty$ for arbitrary initial 
	distributions
	by imposing the ergodicity and the balance condition (BC), 
	which is expressed as $\bm{\pi}=\bm{\pi}\mathsf{T}$.
	In practice, the DBC, 
	which is given by
		\begin{align}
		\pi_i T_{ij}=\pi_j T_{ji},
		\end{align}
	is widely imposed as a sufficient condition of the BC. For instance,   
	one can easily find the transition probability 
	satisfying the DBC such as the Metropolis-Hastings 
	type~\cite{Hastings1970} and the heat-bath type~\cite{Berg1992}. 
	The DBC is also called a reversibility and 
	a Markov chain satisfying the DBC is referred to as 
	a reversible Markov chain.
	In contrast, a Markov chain without the DBC is 
	called an irreversible Markov chain.
        

	\subsection{\label{MHA}Metropolis-Hastings algorithm}
        
	The Metropolis-Hastings algorithm~\cite{Hastings1970}, 
	one of the most famous MCMC algorithms, numerically 
	performs the reversible Markov chain 
	explained in the previous subsection.
	In this subsection, we describe 
	the Metropolis-Hastings algorithm 
	to fix our notation.
       
	First, the transition matrix $\mathsf{T}=(T_{ij})_{i,j\in I}$ 
	satisfying the DBC with respect to $\bm{\pi}$ is decomposed as 
		\begin{align}
		T_{ij}
 		&=
		q_{ij}w_{ij}\quad(i\not=j), \\
		T_{ii}
		&=
		1-\sum_{j\not=i}T_{ij},
		\end{align}
	where $q_{ij}$ and $w_{ij}$ are referred to as the proposal 
	probability and the acceptance probability, respectively.
	$\mathsf{Q}\equiv(q_{ij})_{i,j\in I}$ is a stochastic matrix and 
	$\mathsf{W}\equiv(w_{ij})_{i,j\in I}$ satisfies $0\le w_{ij}\le 1$
	for all $i,j\in I$.
	Let us assume that $\mathsf{Q}$ is symmetric for simplicity.
	$X^{(n)}$ denotes a state after $n$ steps in the Markov chain, 
	starting from an initial state $X^{(0)}$ chosen arbitrarily.
	Then, the elementary procedure of the Metropolis-Hastings 
	algorithm is as follows: 
		\begin{enumerate}[(i)]
		\item Suppose $X^{(n)}=i$. Choose a candidate of 
		new state $j$ according to the distribution given 
		by $\bm{q}_i\equiv(q_{i1},q_{i2},\ldots,q_{i\Omega})$.
		\item Accept the new state $j$ as $X^{(n+1)}=j$ 
		with the probability $w_{ij}$.
		If it is rejected, set $X^{(n+1)}=i$.
		\end{enumerate}

	By repeating the above procedure $M$ times, 
	one can obtain the 
	desired Markov chain $(X^{(n)})_{n=0,1,2,\ldots,M}$ 
	generated by $\mathsf{T}$ and 
	the expectation $\langle\hat{f}\rangle_{\bm{\pi}}$ can be 
	estimated as
		\begin{align}
		\langle\hat{f}\rangle_{\bm{\pi}}\simeq
		\frac1M\sum_{n=1}^{M}f_{X^{(n)}},
		\label{eq:estimator}
		\end{align}
	where $f_{X^{(n)}}$ denotes the value of the quantity $\hat{f}$ 
	after $n$ steps in the Markov chain.
	The convergence rate and the variance of the estimator 
	strongly depend on the target distribution, the choice of 
	the transition matrix, and the quantity to be estimated, 
	although the estimation is proven to be accurate by increasing $M$,
	irrespective of these other factors.


	\subsection{Eigenvalues of the transition matrix and 
	efficiency of MCMC methods}
        
	In this subsection, we survey the relationship between 
	the efficiency of MCMC methods and the eigenvalues 
	of the corresponding transition matrix in the Markov chain. 
        
	\subsubsection{Convergence rate and second-largest eigenvalue}
        
	Let $\mathsf{T}$ be the transition matrix satisfying 
	the DBC with respect to $\bm{\pi}$ and let 
	$\mathsf{B}\equiv{\rm diag}(\pi_1,\ldots,\pi_{\Omega})$ 
	be the diagonal matrix in $\mathbb{R}^{\Omega\times\Omega}$.
	The matrix $\mathsf{B}$ is positive-definite, real-symmetric, 
	and invertible because we have assumed that 
	$\pi_i>0$ for all $i\in I$. 
	Thus, it is well defined that $\mathsf{B}^{1/2}=
	{\rm diag}(\pi_1^{1/2},\ldots,\pi_{\Omega}^{1/2})$ 
	and $\mathsf{B}^{-1/2}=
	{\rm diag}(\pi_1^{-1/2},\ldots,\pi_{\Omega}^{-1/2})$. 

	A similarity transformation of $\mathsf{T}$ with 
	respect to $\mathsf{B}^{-1/2}$ is defined as 
		\begin{align}
		\mathsf{S}\equiv\mathsf{B}^{1/2}\mathsf{T}\mathsf{B}^{-1/2}.
		\end{align}
	The eigenvalues of 
	$\mathsf{S}$ coincide with those of $\mathsf{T}$ including multiplicity 
	because the similarity transformation does not change the 
	characteristic polynomial of $\mathsf{T}$.
	In addition, $\mathsf{S}$ is a real-symmetric matrix and 
	thus all the eigenvalues are ensured to be real 
	because $\mathsf{T}$ is reversible with respect 
	to the target distribution $\bm{\pi}$.
	From the Perron-Frobenius theorem, it is guaranteed that 
	the largest eigenvalue is equal to 1 with multiplicity 1
	and the absolute value of other eigenvalues is less than 1 
	if and only if the transition matrix is ergodic.
	Therefore, when we denote a set of eigenvalues of 
	the reversible transition matrix $\mathsf{T}$ as 
	$\{\lambda_k\}_{k=1}^{K}$,
	the eigenvalues can be rearranged as 
	$1=\lambda_1>\lambda_2\ge\cdots\ge\lambda_{K}>-1$ 
	without loss of generality.
	A set of left eigenvectors $\{\bm{x}_{k,\sigma}\}$ of $\mathsf{S}$ 
	can be set as an orthonormal basis in $\mathbb{R}^{\Omega}$ 
	described as follows:
		\begin{align}
		\bm{x}_{k,\sigma}\mathsf{S}=\lambda_k\bm{x}_{k,\sigma},
		\quad\bm{x}_{k,\sigma}\in\mathbb{R}^{\Omega},\quad
		\bm{x}_{k,\sigma}\bm{x}_{l,\rho}^{\top}=
		\delta_{kl}\delta_{\sigma\rho},
		\end{align}
	where $\sigma\in\{1,\ldots,m_k\}$ is 
	an index for multiplicity with $m_k$ 
	being the multiplicity of $\lambda_k$.
	Left eigenvectors $\{\bm{u}_{k,\sigma}\}$ 
	and right eigenvectors $\{\bm{v}_{k,\sigma}\}$
	of $\mathsf{T}$ are given as 
	$\bm{u}_{k,\sigma}=\bm{x}_{k,\sigma}\mathsf{B}^{1/2}$
	and
	$\bm{v}_{k,\sigma}=\bm{x}_{k,\sigma}\mathsf{B}^{-1/2}$, respectively.
	In particular, the Perron-Frobenius theorem ensures 
	that the left and right eigenvectors 
	associated with 
	$\lambda_1=1$ are given as 
	$\bm{u}_{1,1}=\bm{\pi}$ and 
	$\bm{v}_{1,1}=\bm{1}\equiv(1,\ldots,1)$, respectively.

	Consequently, the reversible transition matrix $\mathsf{T}$ 
	can be decomposed into
		\begin{align}
		\mathsf{T}=\mathsf{A}+\sum_{k=2}^{K}\lambda_k\left(
		\sum_{\sigma=1}^{m_k}
		\bm{v}_{k,\sigma}^{\top}\bm{u}_{k,\sigma}
		\right),
		\label{eq:spectral decomposition}
		\end{align}
	where $\mathsf{A}\equiv\bm{1}^{\top}\bm{\pi}$ is 
	the so-called limiting matrix.
	From the master equation one can derive a formal solution,
		\begin{align}
		\bm{p}^{(n)}=\bm{\pi}+\sum_{k=2}^{K}(\lambda_k)^n
		\left[\sum_{\sigma=1}^{m_k}
		(\bm{p}^{(0)}\bm{v}_{k,\sigma}^{\top})\bm{u}_{k,\sigma}
		\right].
		\label{eq:pn}
		\end{align}
	$\bm{p}^{(n)}$ converges to the stationary distribution 
	$\bm{\pi}$ as $n\to\infty$ because $|\lambda_k|<1$ for 
	$2\le k \le K$.
	Moreover, this indicates that 
	the convergence rate is determined by the second-largest eigenvalue 
	in absolute value, denoted by 
		\begin{align}
		\eta\equiv\max_{2\le k\le K}|\lambda_k|.
		\end{align}
	Thus, the relaxation time, defined by
		\begin{align}
		\tau_{\rm relax}=-\frac1{\log\eta},
		\label{eq:tau_relax}
		\end{align}
	reflects the convergence rate.

	Correspondingly, the expectation value of $\hat{f}$ at time $n$ is 
	also defined as 
		\begin{align}
		\langle\hat{f}\rangle_n\equiv\sum_{i=1}^{\Omega}
		p_i^{(n)}f_i=\bm{p}^{(n)}\bm{f}^{\top}.
		\end{align}
	From Eq.~(\ref{eq:pn}), it can be rewritten as
		\begin{align}
		\langle\hat{f}\rangle_n=\langle\hat{f}\rangle_{\bm{\pi}}+
		\sum_{k=2}^{K}(\lambda_k)^n\left[\sum_{\sigma=1}^{m_k}
		(\bm{p}^{(0)}\bm{v}_{k,\sigma}^{\top})
		(\bm{u}_{k,\sigma}\bm{f}^{\top})\right],
		\end{align}
	which may indicate that the convergence rate of the expectation is 
	also determined by $\eta$. 
	However, the eigenvalue 
	$\lambda_k$ does not affect the relaxation dynamics of 
	$\langle\hat{f}\rangle_n$ 
	if $\bm{u}_{k,\sigma}\bm{f}^{\top}=0$ holds for all 
	$\sigma=1,\ldots,m_k$. 
	In particular, the convergence rate of 
	$\langle\hat{f}\rangle_n$ is not determined by 
	$\lambda_2$ if $\bm{f}$ is orthogonal to 
	all the left eigenvectors associated with 
	the second-largest eigenvalue.

                
	\subsubsection{Correlations and second-largest eigenvalue}

	If the central limit theorem holds, an effective variance of 
	the estimator in Eq.~(\ref{eq:estimator}) is given by 
	$v(\hat{f},\bm{\pi},\mathsf{T})/M$ for sufficiently large $M$,
	where $v(\hat{f},\bm{\pi},\mathsf{T})$
	denotes the asymptotic variance defined by  
		\begin{align}
		v(\hat{f},\bm{\pi},\mathsf{T})\equiv
		\lim_{M\to\infty}M{\rm var}\left[
		\frac1M\sum_{n=1}^{M}f_{X^{(n)}}\right].
		\end{align}
	The asymptotic variance is often enlarged by the correlation 
	between samples in the Markov chain. 
	It is convenient to measure the correlation by an integrated 
	autocorrelation time, defined as
		\begin{align}
		\tau_{{\rm int},\hat{f}}
		=\sum_{n=1}^{\infty}C^{(n)}_{\hat{f}},
		\end{align}
	where   
		\begin{align}
		C^{(n)}_{\hat{f}}\equiv
		\frac{\langle\hat{f}_{\bm{\pi}}\hat{f}\rangle_n
		-\langle\hat{f}\rangle_{\bm{\pi}}^2}{
		{\rm var}_{\bm{\pi}}[\hat{f}]}
		\end{align}
	denotes the autocorrelation function, ${\rm var}_{\bm{\pi}}[\hat{f}]
	\equiv\langle\hat{f}^2\rangle_{\bm{\pi}}
	-\langle\hat{f}\rangle_{\bm{\pi}}^2$ is 
	the variance of $\hat{f}$ for an independent sampling, and 
	$\langle\hat{f}_{\bm{\pi}}\hat{f}\rangle_n\equiv
	\bm{f}\mathsf{B}\mathsf{T}^n\bm{f}^{\top}$.
	The relationship between the asymptotic variance and the integrated  
	autocorrelation time is explicitly given by    
		\begin{align}
		v(\hat{f},\bm{\pi},\mathsf{T})=
		(1+2\tau_{{\rm int}, \hat{f}}){\rm var}_{\bm{\pi}}[\hat{f}],
		\end{align}
	indicating that strong correlation results in poor estimation.

	Here, we define the fundamental matrix $\mathsf{Z}$ as
		\begin{align}
		\mathsf{Z}\equiv(\mathsf{I-T+A})^{-1}=\mathsf{I}+
		\sum_{n=1}^{\infty}(\mathsf{T}^n-\mathsf{A}).
		\end{align}
	Using the matrix $\mathsf{Z}$,
	we can rewrite the asymptotic variance as~\cite{Bremaud}
		\begin{align}
		v(\hat{f},\bm{\pi},\mathsf{T})
		&=
		\bm{f}[\mathsf{BZ}+(\mathsf{BZ})^{\top}
		-\mathsf{B-BA}]\bm{f}^{\top} \notag \\
		&=
		\sum_{k=2}^{K}\frac{1+\lambda_k}{1-\lambda_k}
		\left[\sum_{\sigma=1}^{m_k}
		(\bm{f}\mathsf{B}\bm{v}_{k,\sigma}^{\top})
		(\bm{u}_{k,\sigma}\bm{f}^{\top})\right].
		\end{align}
	It turns out that the asymptotic variance 
	depends on all the eigenvalues of the transition matrix.                
	However, the upper bound of the ratio between the variance and 
	the asymptotic variance is determined only by 
	the second-largest eigenvalue
	of the transition matrix as
		\begin{align}
		\max_{\hat{f}\not=\hat{0}}
		\frac{v(\hat{f},\bm{\pi},\mathsf{T})}{
		{\rm var}_{\bm{\pi}}[\hat{f}]}=
		\frac{1+\lambda_2}{1-\lambda_2},
		\end{align}
	where the equality is attained by $\bm{f}$ given by 
	the linear combination of $\{\bm{v}_{2,\sigma}\}_{\sigma=1}^{m_2}$.


	\subsubsection{Irreversible transition matrix and 
	additive reversibilization} 

	A transition matrix that has a unique stationary distribution 
	but does not satisfy the DBC 
	with respect to the stationary distribution 
	is referred to as an irreversible transition matrix.
	Let us consider the irreversible transition matrix 
	$\mathsf{T}$ that is ergodic and has a unique 
 	stationary distribution $\bm{\pi}$.
	Although some eigenvalues might be complex and 
	it is not guaranteed that $\mathsf{T}$ is diagonalizable in general,
	the transition matrix $\mathsf{T}$ is diagonalizable 
	if all the eigenvalues of $\mathsf{T}$ are distinct. 
	In this case, one can rearrange 
	the eigenvalues of $\mathsf{T}$ as $1=\lambda_1>|\lambda_2|\ge\cdots\ge
	|\lambda_{\Omega}|\ge 0$ and impose a normalization 
	condition $\bm{u}_{k}\bm{v}_k^{\top}=1$
	without loss of generality.
	Thus, the formula in Eq.~(\ref{eq:spectral decomposition}) 
	also holds in an irreversible Markov chain.
	This implies that the convergence rate
	toward the stationary distribution is determined by $|\lambda_2|$
	and the asymptotic variance depends on all the eigenvalues 
	even in an irreversible case.

	Let us introduce the symmetrized transition matrix of
	the irreversible transition matrix $\mathsf{T}$, 
	defined as follows~\cite{Bremaud}:
		\begin{align}
		\mathsf{T}_0\equiv\frac12(\mathsf{T}
		+\mathsf{B}^{-1}\mathsf{T}\mathsf{B});
		\end{align}
	this is referred to as the additive reversibilization. 
	Note that the additive reversibilization $\mathsf{T}_0$ 
	satisfies the DBC with respect to $\bm{\pi}$.
	In addition, it is theoretically proved that 
	the irreversible transition matrix is always better than
	that of the corresponding additive reversibilization in terms of 
	the real part of the second-largest eigenvalue~\cite{Ichiki2013} and 
	the asymptotic variance~\cite{Sun2010}.
        
	When the irreversible transition matrix $\mathsf{T}$ 
	is constructed by modifying a reversible transition matrix 
	$\mathsf{T}_{\mathrm{rev}}$, 
	we are interested in comparing
	the efficiency of $\mathsf{T}$ with 
	that of $\mathsf{T}_{\mathrm{rev}}$.
	Although the relation between the efficiency of $\mathsf{T}$ and 
	that of the additive reversibilization of $\mathsf{T}$ is
	discussed in Refs.~\cite{Ichiki2013, Sun2010}, 
	the additive reversibilization of $\mathsf{T}$ 
 	is not always equivalent to $\mathsf{T}_{\mathrm{rev}}$ and 
	the relation between 
	the efficiency of $\mathsf{T}$ and that of $\mathsf{T}_{\mathrm{rev}}$
	in general is not yet well understood.
	In Sec.~\ref{Sect:Performance Evaluation}, the relation is
	discussed by the eigenvalue analysis of the transition matrix
	for a specific probabilistic model. 


	\section{\label{Sect:MCwithSDBC}
	Markov Chain with the Skew Detailed Balance Condition}
        
	The Markov chain with the skew detailed balance condition 
	has been proposed by Turitsyn et al.~\cite{Turitsyn2011}. 
	In this section, we review how to construct an irreversible 
	Markov chain by imposing the SDBC in general. 

        
	\subsection{\label{subsect:Extension of s.s.}
	Extension of the stationary distribution}
        
	First, we double the state space $I$ 
	by introducing an auxiliary variable 
 	$\varepsilon\in\{+,-\}$. The extended state space is 
	given as $\tilde{I}:=I\times\{+,-\}$ and 
	a state in $\tilde{I}$ is described by $(i,\varepsilon)$.
	Let $\pi_{(i,\varepsilon)}$ be the probability of finding the 
	state $(i,\varepsilon)$ and $\pi_{(i,\varepsilon)}$ be uniform 
	with respect to $\varepsilon$. Then, the target
	distribution is extended to 
		\begin{align}
		\tilde{\bm{\pi}}
		&\equiv(\pi_{(1,+)},\ldots,
		\pi_{(\Omega,+)}, 
		\pi_{(1,-)},\ldots,\pi_{(\Omega,-)}) \notag \\
		&=
		\frac12(\bm{\pi},\bm{\pi}).
		\end{align}
        
	Let $\hat{f}$ be a quantity defined on the extended state space. 
	$f_{(i,\varepsilon)}$ denotes the realization of $\hat{f}$ 
	at the extended state $(i,\varepsilon)$.
	Then, the expectation value of $\hat{f}$ with respect to 
	the extended target distribution is defined as
		\begin{align}
		\langle\hat{f}\rangle_{\tilde{\bm{\pi}}}\equiv
 		\sum_{i=1}^{\Omega}\sum_{\varepsilon=\pm}
		\pi_{(i,\varepsilon)}f_{(i,\varepsilon)}.
		\end{align}
	In the case in which $\hat{f}$ is independent of $\varepsilon$, 
	{\rm i.e.}, $f_{(i,\varepsilon)}=f_i$, 
 		\begin{align}
		\langle\hat{f}\rangle_{\tilde{\bm{\pi}}}=
		\sum_{i=1}^{\Omega}\sum_{\varepsilon=\pm}\frac{\pi_i}2 f_i=
		\sum_{i=1}^{\Omega}\pi_i f_i=
		\langle\hat{f}\rangle_{\bm{\pi}}.
		\end{align}
	In other words, 
	the expectation with respect to the extended target distribution 
	corresponds to that with respect to the original target distribution.


	\subsection{\label{subsect:SDBC}Skew detailed balance condition}

\begin{figure}[tb]
		\centering
		\includegraphics[width=.47\textwidth]{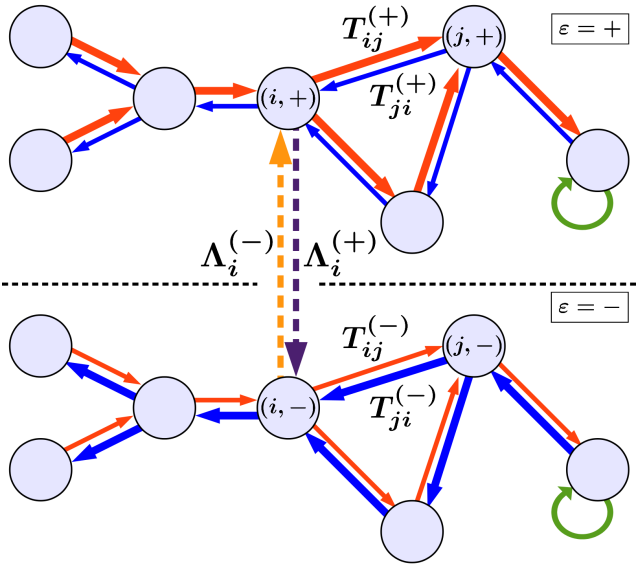}
		\caption{(Color online) Schematic picture of 
		the extended transition graph.}
		\label{fig:extended-statespace}
		\end{figure}

	Here, we construct the Markov chain on the extended 
	state space $\tilde{I}$.
	The transition matrix in the Markov chain on $\tilde{I}$
	is given as follows:
		\begin{align}
		\tilde{\mathsf{T}}=
		\left(
			\begin{matrix}
			\mathsf{T}^{(+)} & \mathsf{\Lambda}^{(+)} \\
			\mathsf{\Lambda}^{(-)} & \mathsf{T}^{(-)} 
			\end{matrix}
		\right),
	\end{align}
	where $\mathsf{T}^{(\pm)}=(T^{(\pm)}_{ij})_{i,j\in I}$ 
	and $\mathsf{\Lambda}^{(\pm)}=
	{\rm diag}(\Lambda^{(\pm)}_i)_{i\in I}$.
	$T^{(\varepsilon)}_{ij}$ denotes the transition probability
	from state $(i,\varepsilon)$ to state $(j,\varepsilon)$ and 
	$\Lambda^{(\varepsilon)}_i$ is that 
	from state $(i,\varepsilon)$ 
	to state $(i,-\varepsilon)$ (Fig.~\ref{fig:extended-statespace}).
	Conservation of probability in the transition matrix is
	expressed as 
		\begin{align}
		\sum_{j\in I}T_{ij}^{(\pm)}+\Lambda_i^{(\pm)}=1,
		\label{eq:conservation of probability}
		\end{align}
	for all $i\in I$. 
        
	Let us assume that $\tilde{\mathsf{T}}$ is ergodic. 
	Then, the BC 
	$\tilde{\bm{\pi}}=\tilde{\bm{\pi}}\tilde{\mathsf{T}}$
	ensures that the stationary distribution 
	of $\tilde{\mathsf{T}}$ is exactly $\tilde{\bm{\pi}}$. 
	To construct an irreversible Markov chain,  
	we impose the SDBC~\cite{Turitsyn2011} 
	described as
		\begin{align}
		\pi_i T^{(+)}_{ij} = \pi_j T^{(-)}_{ji}.
		\label{eq:SDBC}
		\end{align}
	This condition means that the stochastic flow with a transition 
	from state $(i,+)$ to state $(j,+)$ balances with the transition 
	from state $(j,-)$ to state $(i,-)$. 
	In general, the DBC is violated by imposing the SDBC.
                
	By the conservation of probability in Eq.~(\ref{eq:conservation 
	of probability}) and the SDBC in Eq.~(\ref{eq:SDBC}), the 
	BC can be rewritten as 
		\begin{align}
		\Lambda^{(+)}_i-\Lambda^{(-)}_i=\sum_{j\not=i}
		(T^{(-)}_{ij}-T^{(+)}_{ij}).
		\label{eq:SBC2}
		\end{align}
	Consequently, the convergence 
	to the extended stationary distribution 
	is guaranteed by imposing the SDBC and Eq.~(\ref{eq:SBC2}).         
        

	\section{Irreversible Metropolis-Hastings Algorithm}
	\label{Sect:IMH}

	In this section, we construct the MCMC method on the basis of 
	the SDBC. Although the prototype algorithm for 
	a mean-field Ising model
	has been proposed by Turitsyn et al.~\cite{Turitsyn2011}, 
	we develop the algorithm so as to be applicable 
	to more general systems. 


	\subsection{\label{subsect:Choice of T}Choice of the transition matrix} 

	An example of the transition probability 
	$T_{ij}^{(\pm)}$ is given in this subsection.
	To begin, we prepare a transition matrix on $I$ satisfying the DBC 
	with respect to $\bm{\pi}$. Namely, 
	a transition matrix $\mathsf{T}=(T_{ij})_{i,j\in I}$ 
	with $\pi_i T_{ij} = \pi_j T_{ji}$ is given.
	The transition probability $T^{(\pm)}_{ij}$
	satisfying the SDBC in Eq.~(\ref{eq:SDBC}) is obtained by 
	modifying $T_{ij}$ with an arbitrary function $\Delta_{ij}^{(\pm)}$ 
	of two states $i,j\in I$ as follows:
		\begin{equation}
		T^{(\pm)}_{ij}=
		\frac{1+\Delta_{ij}^{(\pm)}}{2}T_{ij},
		\label{eq:stp}
		\end{equation}
	where $\Delta_{ij}^{(\pm)}$ satisfies $|\Delta_{ij}^{(\pm)}|\le 1$ 
	and $\Delta_{ij}^{(+)}=\Delta_{ji}^{(-)}$
	for all $i,j\in I$. It is straightforward to show that 
	$T^{(\pm)}_{ij}$ satisfies the SDBC in Eq.~(\ref{eq:SDBC}).

	Even if the transition probability $T_{ij}^{(\pm)}$ is 
	fixed as Eq.~(\ref{eq:stp}), 
	there remain several choices of the transition probability 
	$\Lambda_i^{(\pm)}$. 
	The following transition probabilities 
	satisfy the condition of Eq.~(\ref{eq:SBC2}): 
		\begin{align}
		\Lambda_{i,{\rm SH_1}}^{(\pm)}=
		\sum_{j\not=i}T_{ij}^{(\mp)}=
 		\sum_{j\not=i}\frac{1+\Delta_{ij}^{(\mp)}}{2}T_{ij},
		\label{eq:SH1}
		\end{align}
		\begin{align}
		\Lambda_{i,{\rm SH_2}}^{(\pm)}=
		\frac12\mp\frac14\sum_{j\not=i}
		(\Delta_{ij}^{(+)}-\Delta_{ij}^{(-)})T_{ij},
		\label{eq:SH2}          
		\end{align}
		\begin{align}
		\Lambda_{i,{\rm SH_3}}^{(\pm)}=
		\sum_{j\not=i}\frac{1-\Delta_{ij}^{(\pm)}}{2}T_{ij},
		\label{eq:SH3}
		\end{align}
	and
		\begin{align}
		\Lambda_{i,{\rm TCV}}^{(\pm)}
		&=
		\max\Big\{0,~\sum_{j\not=i}(T_{ij}^{(\mp)}-
		T_{ij}^{(\pm)})\Big\} \notag \\
		&=
		\max\Big\{0,~\mp\frac12\sum_{j\not=i}(\Delta_{ij}^{(+)}
		-\Delta_{ij}^{(-)})T_{ij}\Big\}. 
		\end{align}
	They are referred to as the Sakai-Hukushima 1 (SH$_1$) type, 
	the Sakai-Hukushima 2 (SH$_2$) type~\cite{YS2013}, 
	the Sakai-Hukushima 3 (SH$_3$) type, and
	the Turitsyn-Chertkov-Vucelja (TCV) type~\cite{Turitsyn2011}, 
	respectively.
        
        
	\subsection{\label{subsect:IMH}
	Irreversible Metropolis-Hastings algorithm}

	Let us decompose the transition probability $T_{ij}$ as 
	$T_{ij}=q_{ij}w_{ij}$ $(i\not=j)$ as was done in the previous section.
	$\tilde{X}^{(n)}$ denotes the state in $\tilde{I}$ after $n$ 
	steps in a Markov chain.  
	Then, the irreversible Metropolis-Hastings (IMH) algorithm is 
	described as follows:
        
		\begin{enumerate}[(i)]
        
		\item Set an initial state $\tilde{X}^{(0)}$ chosen arbitrarily.
        
 		\item Suppose that $\tilde{X}^{(n)}=(i,\varepsilon)$ 
		and propose a new state $(j,\varepsilon)$ 
		by using the probability distribution $(q_{ij})_{j\in I}$.
        
		\item Accept the proposed state as 
		$\tilde{X}^{(n+1)}=(j,\varepsilon)$ 
		with the probability $(1+\Delta_{ij}^{(\varepsilon)})w_{ij}/2$.

		\item If the proposed state is rejected in step (iii), 
		set $\tilde{X}^{(n+1)}=(i,-\varepsilon)$ with the probability 
		$p$ given as
			\begin{align}
			p = \frac{\Lambda_i^{(\varepsilon)}}{\dis
			1-\sum_{j\not=i}T_{ij}^{(\varepsilon)}}.
			\end{align}
		If also rejected, set $\tilde{X}^{(n+1)}=\tilde{X}^{(n)}$.
		\end{enumerate}
        
		By repeating the above procedures (ii)--(iv) $M$ times, 
		one can obtain the Markov chain 
		$(\tilde{X}^{(n)})_{n=0,1,2,\ldots,M}$ 
		generated by the transition matrix $\tilde{\mathsf{T}}$, 
		verified in Appendix~\ref{sect:A}.
		The expectation $\langle\hat{f}\rangle_{\tilde{\bm{\pi}}}$ is 
		estimated in the same way as in the original Metropolis-Hastings
		algorithm. 
        
                
	\section{\label{Sect:Performance Evaluation} Performance Evaluation}

	In this section, 
	we discuss a random walk on a circle as a toy model. 
	By specifying all the eigenvalues and eigenvectors of the 
	corresponding transition matrices, we discuss the efficiency
	of the irreversible MCMC method we have proposed 
	in the previous sections.
        
	\subsection{\label{subsect:Random walk on circle} 
	Random walk on a circle}

\begin{figure}[tb]
		\centering
		\includegraphics[width=.47\textwidth]{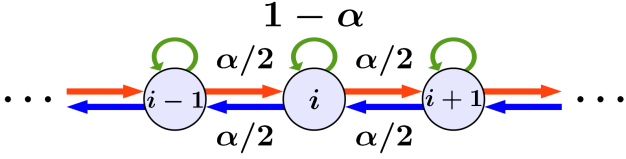}
		\caption{(Color online) Schematic picture of 
		a part of the transition graph on the state space $I$ 
		in a reversible Markov chain.}
		\label{fig:1dRW}
		\end{figure}

	Suppose that there are states $i=1,2,\ldots,\Omega$ on a circle
	under a periodic boundary condition.
	Then, we give the transition matrix on the state space as 
		\begin{align}
		\mathsf{T}=(1-\alpha)\mathsf{I}_{\Omega}
		+\mathsf{J}_{\Omega}\left(\frac{\alpha}2,\frac{\alpha}2\right),
		\end{align}
	where $0<\alpha<1$ is a transition rate, 
	$\mathsf{I}_{\Omega}$ is the $\Omega$th identity matrix, and 
		\begin{align}
		\mathsf{J}_{\Omega}(a,b)\equiv\left(
			\begin{matrix}
			0 & a &   &   & b \\	
			b & 0 & a &   &   \\
			& \ddots & \ddots & \ddots & \\
			  &   & b & 0 & a \\
			a &   &   & b & 0 
			\end{matrix}
			\right),
		\end{align}
	respectively. Figure~\ref{fig:1dRW} illustrates 
	a transition graph on $I$.
	The stationary distribution of $\mathsf{T}$ is given by 
	the uniform distribution as 
	$\pi_i=1/\Omega$ since the transition matrix is doubly stochastic.
	Notice that the transition matrix satisfies the DBC with 
	respect to the uniform distribution.
        
	All the eigenvalues of the transition matrix for the toy model 
	are derived as 
		\begin{align}
		\lambda_k=1-\alpha+\alpha\cos\theta_k,
		\label{eqn:eigenvalue_DBC}
		\end{align}
	where $\theta_k\equiv2\pi(k-1)/\Omega$ for $1\le k\le[\Omega/2]+1$
	and $[x]$ denotes the maximum integer that does not exceed 
	$x\in\mathbb{R}$. 
	The multiplicities of eigenvalues for even $\Omega$ are given by 
	$m_1=1$, $m_k=2$ for $2\le k\le [(\Omega+1)/2]$, 
	and $m_{\Omega/2+1}=1$, 
	respectively. The eigenvectors of $\mathsf{T}$ are the same as those of 
	$\mathsf{J}_{\Omega}(\alpha/2,\alpha/2)$, 
	described in Appendix~\ref{sect:B} in detail.

                
	\subsection{\label{subsect:irreversible 1dRW}
	Irreversible random walk with the SDBC}
        
	Some previous works addressed 
	the effectiveness of the violation of the DBC. 
	Diaconis et al. have analyzed the convergence rate of 
	the total variation and $\chi^2$ distance of 
	a nonreversible random walk~\cite{Diaconis1997}. 
	Chen et al. have shown          
	that the mixing time of the random walk 
	is reduced by violating the DBC~\cite{Chen1999}.
	In this subsection, 
	we derive all the eigenvalues and eigenvectors of 
	the extended transition 
	matrix with the SDBC. By using them, we reveal that 
	the convergence rate and 
	the worst evaluation of the asymptotic variance are 
	improved by imposing the SDBC.
	We also find that the violation of the DBC is not always superior to
	the original method with the DBC. 
        
\begin{figure}[tb]
		\centering
		\includegraphics[width=.47\textwidth]{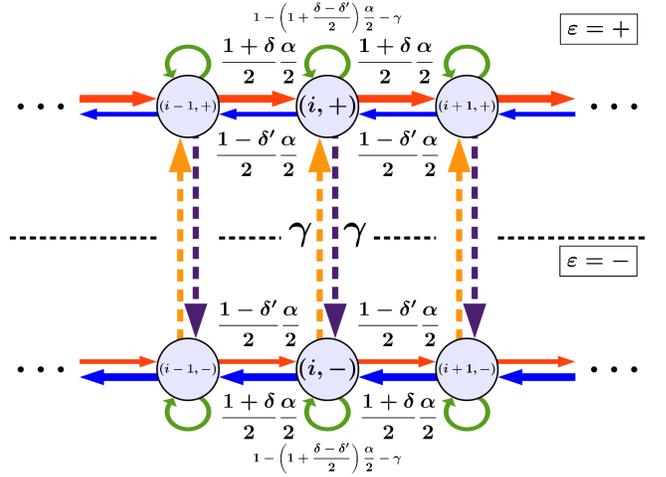}
		\caption{(Color online) Schematic picture of a part of 
		the transition graph on the extended state space in an
		irreversible Markov chain.}
		\label{fig:irreversible-1dRW}
		\end{figure}

	Let us apply the methodology of the SDBC to 
	the random walk described in the previous subsection. 
	By adding the auxiliary variable $\varepsilon\in\{+,-\}$, 
	the state space is doubled and the target distribution is 
	extended as the uniform distribution on the extended state space.
	The transition matrix on the extended state space is given as
		\begin{align}
		\tilde{\mathsf{T}}
		=&
		\left[1-\left(1+\frac{\delta-\delta'}2\right)
		\frac{\alpha}2-\gamma\right]\mathsf{I}_{2\Omega}
		\notag \\
		&\quad +\tilde{\mathsf{J}}_{2\Omega}\left(
		\frac{1+\delta}2\frac{\alpha}2,\frac{1-\delta'}2
		\frac{\alpha}2;\gamma
		\right),
		\end{align}
	where 
		\begin{align}
		\tilde{\mathsf{J}}_{2\Omega}(a,b;c)\equiv\left(
			\begin{matrix}
			\mathsf{J}_{\Omega}(a,b) & c\mathsf{I}_{\Omega} \\
			c\mathsf{I}_{\Omega} & \mathsf{J}_{\Omega}(b,a)
			\end{matrix}
		\right),        
		\end{align}
	with $|\delta|\le 1$ and $|\delta'|\le 1$ being parameters to control 
	the violation of the DBC, 
	and where $\gamma$ 
	is a transition rate for $\varepsilon$ flip.
	The allowed range of $\gamma$, 
	$0\le\gamma<1-(2+\delta-\delta')\alpha/4$,
	includes the particular transition rates such as 
	SH$_1$, SH$_2$, SH$_3$, and TCV types,
	explicitly specified by 
	$\gamma_{\rm SH_1}=(2+\delta-\delta')\alpha/4$, 
	$\gamma_{\rm SH_2}=1/2$, 
	$\gamma_{\rm SH_3}=(2-\delta+\delta')\alpha/4$, and 
	$\gamma_{\rm TCV}=0$, respectively.
	The transition graph on the extended state space is given 
	as Fig.~\ref{fig:irreversible-1dRW}.

	All the eigenvalues of $\tilde{\mathsf{T}}$ are obtained with
	Eq.~(\ref{eqn:eigenvalue_DBC}) as 
		\begin{align}
		\tilde{\lambda}_k^{\pm}=&
		1-\left(1+\frac{\delta-\delta'}2\right)
		\alpha\sin^2\frac{\theta_k}2-\gamma
		\notag \\
		&\quad
		\pm\sqrt{\gamma^2-\left(\frac{\delta+\delta'}2\right)^2\left(
		\frac{\alpha}2\sin\theta_k\right)^2},
		\label{eq:eigenvalues-tildeT}
		\end{align}
	for $1\le k\le[\Omega/2]+1$.
	The multiplicities of eigenvalues, depending only on the label $k$, 
	are given by $m_1=1$, 
	$m_k=2$ for $2\le k\le[(\Omega+1)/2]$, 
	and $m_{\Omega/2+1}=1$ if $\Omega$ is even, respectively.
	We should note that some eigenvalues are degenerate or
	might be complex depending on $\delta$, $\delta'$, and $\gamma$.
	In particular, the extended transition matrix $\tilde{\mathsf{T}}$ is 
	not diagonalizable when 
	$\gamma=|\delta+\delta'|(\alpha/4)\sin\theta_k$ for 
	some $2\le k\le [(\Omega+1)/2]$.
	All the eigenvectors of the extended transition matrix 
	$\tilde{\mathsf{T}}$ are the same as 
	those of $\tilde{J}_{2\Omega}(a,b;c)$ 
	with $a=(1+\delta)\alpha/4$, $b=(1-\delta')\alpha/4$, and $c=\gamma$,
	as described in Appendix~\ref{sect:C} in detail.


	\subsection{Comparison of efficiency}
        
	In the present random-walk problem, 
	the extended transition matrix $\tilde{\mathsf{T}}$ with 
	$\delta=\delta'=0$ is equivalent to the additive reversibilization of 
	$\tilde{\mathsf{T}}$ with $\delta=\delta'(\not=0)$.
	Thus, it is theoretically guaranteed 
	from Refs.~\cite{Ichiki2013, Sun2010} that 
	the real part of the second-largest eigenvalue and the asymptotic 
	variance of $\tilde{\mathsf{T}}$ with $\delta=\delta'(\not=0)$ 
	are reduced in comparison with those of $\tilde{\mathsf{T}}$ 
	with $\delta=\delta'=0$.
	However, it is unclear whether the convergence rate and 
	the asymptotic variance of $\tilde{\mathsf{T}}$ are 
	reduced in comparison with those of the original reversible
	transition matrix $\mathsf{T}$.
	In this subsection, we study two particular cases: 
	(A) $\delta=\delta'$ and (B) $\delta'=1$. 
	Note that in the case (B), a path from state $(i,\varepsilon)$ to 
	$(i-\varepsilon,\varepsilon)$ in the corresponding transition graph 
	vanishes for all $i$.
	We assume that $\alpha=1/2$ and that 
	$\Omega$ is a sufficiently large even number for simplicity. 
	By using the explicit expression of eigenvalues and eigenvectors                 
	derived in the previous subsections and appendices, 
	we show analytically that the irreversible Markov chain 
	with the SDBC for 
	the random walk is more efficient than the reversible one 
	in terms of the relaxation time and the asymptotic variance.

                        
	\subsubsection{Convergence rate}
        
	First, let us discuss the relaxation time of the random-walk problem.  
	In the case of the reversible random walk in
	Sec.~\ref{subsect:Random walk on circle}, 
	the relaxation time is obtained as
		\begin{align}
		\tau_{\rm relax}=-\frac1{\log{|\lambda_2|}}.
		\end{align}
	In contrast, in the irreversible random walk of cases (A) and (B), 
	the relaxation time is obtained as 
		\begin{align}
		\tilde{\tau}_{\rm relax}=-\frac1{\log{\tilde{\eta}}},
		\end{align}
	where 
		\begin{align}
		\tilde{\eta}\equiv\max_{\substack{
		2\le k\le \Omega/2+1 \\ \varepsilon=\pm}}
		|\tilde{\lambda}_k^{\varepsilon}|
		\end{align}
	is the second-largest eigenvalue in absolute value and
	the candidates of $\tilde{\eta}$ are $|\tilde{\lambda}_1^-|$, 
	$|\tilde{\lambda}_2^+|$, and $|\tilde{\lambda}_{\Omega/2+1}^-|$.
	Then, $\tilde{\eta}$ is identified as follows:
		\begin{widetext}                        
			\begin{align}
			\left\{
				\begin{array}{ll}
				\tilde{\eta}=|\tilde{\lambda}_1^-| & \text{if}\quad
				0\le\gamma\le\min\left[ 
				\frac14\left(\sin^2\frac{\pi}{\Omega}
				+\delta^2\cos^2\frac{\pi}{\Omega}\right)\right.,\\
				&\qquad\left.\frac14\left(1-\frac12\sin^2\frac{\pi}{\Omega}\right)
				-\frac14\sqrt{\left(1-\frac12\sin^2\frac{\pi}{\Omega}\right)
				\left(1-\frac52\sin^2\frac{\pi}{\Omega}\right)
				+\frac{\delta^2}4\sin^2\frac{2\pi}{\Omega}}\right], \\
				\tilde{\eta}=|\tilde{\lambda}_{\Omega/2+1}^-| & \text{if}\quad
				\frac9{16}\left(1-\frac13\sin^2\frac{\pi}{\Omega}\right)
				-\frac3{16}\sqrt{\left(1-\frac13\sin^2\frac{\pi}{\Omega}\right)^2
				-\frac{\delta^2}{18}\sin^2\frac{2\pi}{\Omega}}
				\le\gamma\le\frac34, \\
				\tilde{\eta}=|\tilde{\lambda}_2^+| & \text{otherwise}
				\end{array}
			\right.
			\end{align}
		for case (A) and                    
			\begin{align}
			\left\{
				\begin{array}{ll}
				\tilde{\eta}=|\tilde{\lambda}_1^-| & \text{if}\quad
				0\le\gamma\le\frac14\left(1+\frac{1+\delta}4
				\sin^2\frac{\pi}{\Omega}\right)-\frac14\sqrt{
				\left(1+\frac{1+\delta}4\sin^2\frac{\pi}{\Omega}\right)^2
				-\frac{(1+\delta)(7-\delta)}4\sin^2\frac{\pi}{\Omega}},\\
				\tilde{\eta}=|\tilde{\lambda}_{\Omega/2+1}^-| & \text{if}\quad
				\frac34\left[1-\frac{1+\delta}8
				\left(1+\sin^2\frac{\pi}{\Omega}\right)\right]+\frac14\sqrt{
				\left[1-\frac{1+\delta}8\left(
				1+\sin^2\frac{\pi}{\Omega}\right)\right]^2
				-\frac{(1+\delta)^2}{32}\sin^2\frac{2\pi}{\Omega}}\le\gamma\le
				\frac{7-\delta}8,\\
				\tilde{\eta}=|\tilde{\lambda}_2^+| & \text{otherwise}
				\end{array}
			\right.
			\end{align}
		for case (B), respectively. 
			\begin{figure*}[t]
			\centering
			\includegraphics[width=.95\textwidth]{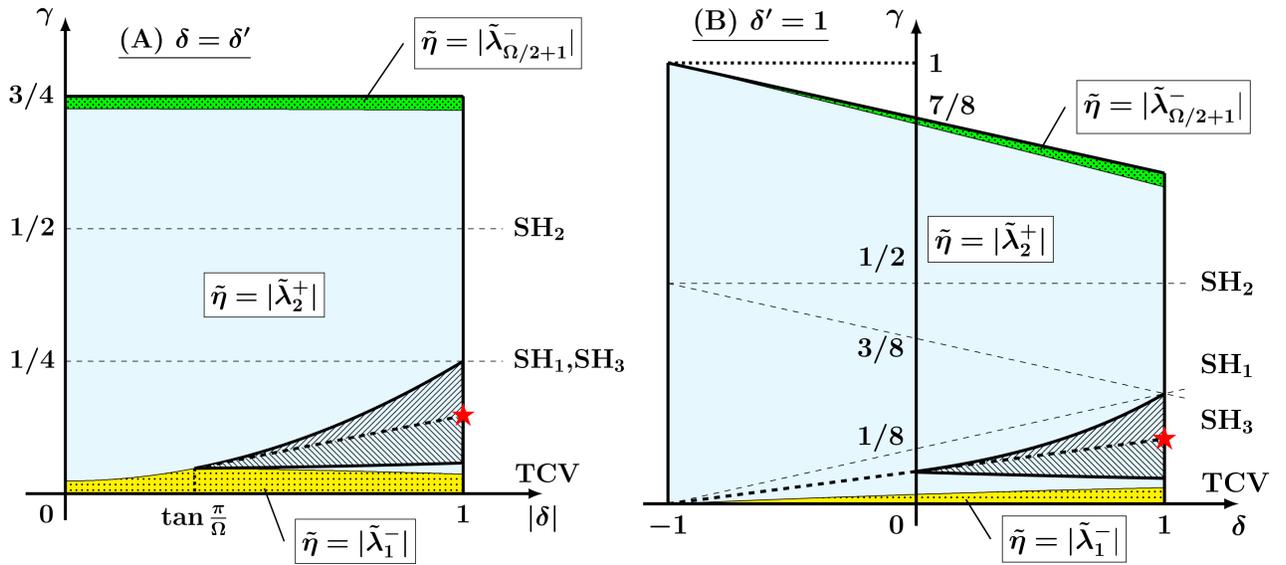}        
 			\caption{(Color online) Maps of the second-largest eigenvalue of the 
			extended transition matrix $\tilde{\mathsf{T}}$ in absolute value.
			The left and right panels represent cases (A) $\delta=\delta'$ and 
			(B) $\delta'=1$, respectively.
			In each panel, the shaded portion represents the 
			parameter region $(\delta,\gamma)$ where the relaxation time is 
			reduced by imposing the skew detailed balance condition.
			The thick dashed line represents the minimum of the relaxation time 
			of the irreversible random walk 
			for a fixed $\delta$ and the star symbol provides
			the lowest value of the relaxation time in the parameter region.
			Note that $\tilde{\lambda}_2^{+}$ is a 
			complex number below the thick dashed line.}
			\label{fig:SLEM}
			\end{figure*}                   
		By comparing $\tau_{\rm relax}$ with $\tilde{\tau}_{\rm relax}$, 
		we find that $\tau_{\rm relax}\ge\tilde{\tau}_{\rm relax}$
		in the parameter region of $(\delta,\gamma)$, explicitly given by 
			\begin{align}
			\frac12\sin^2\frac{\pi}{\Omega}
			+\frac1{16}\frac{\sin^2\frac{2\pi}{\Omega}}{
			1+\cos^2\frac{\pi}{\Omega}}
			\left(\delta^2-\tan^2\frac{\pi}{\Omega}\right)
			\le\gamma\le\frac14\left(
			\sin^2\frac{\pi}{\Omega}+\delta^2\cos^2\frac{\pi}{\Omega}\right)
			\end{align}
		for case (A) and 
			\begin{align}
			\frac{4\sin^2\frac{\pi}{\Omega}}{4-(1+\delta)\sin^2\frac{\pi}{\Omega}}
			\left[\cos^2\frac{\pi}{\Omega}+\left(\frac{3-\delta}4\right)^2\right]
			\le\gamma\le\frac{3-\delta}8\left[\sin^2\frac{\pi}{\Omega}
			+\left(\frac{1+\delta}{3-\delta}\right)^2
			\cos^2\frac{\pi}{\Omega}\right]
			\end{align}
		for case (B), respectively. 
		The parameter region of $(\delta, \gamma)$ is shown 
		by the shaded areas in Fig.~\ref{fig:SLEM}.
		\end{widetext}

	In particular, $\tilde{\tau}_{\rm relax}$ is minimized 
	by setting $|\delta|=1$ and 
	$\gamma=(1/4)\sin(2\pi/\Omega)$ in both cases.
	This is the best choice of 
	parameters in terms of the relaxation time. 
	Then, the $\Omega$ dependence of the relaxation time is 
	qualitatively improved from $O(\Omega^2)$ in the reversible case 
	to $O(\Omega)$ in the irreversible case, 
	meaning that the violation of the DBC yields the qualitative change of 
	the relaxation dynamics in the random walk from diffusive to ballistic.
	However, the relaxation time increases from the minimum value 
	if the transition rate $\gamma$ is set as the SH$_1$, SH$_2$, SH$_3$, 
	and TCV types previously attained~\cite{Turitsyn2011, YS2013}. 
	This result implies that the efficiency is not always improved 
	even if the DBC is violated.

	One may consider that the irreversible Markov chains can possibly have
	complex eigenvalues.
	In fact, as shown in Fig.~\ref{fig:SLEM}, there is a finite region where
	the second-largest eigenvalue is complex. Interestingly, the parameter
	set with the minimum relaxation time is located on the boundary of the
	region. However, the emergence of the complex eigenvalues does not
	always involve the efficiency of the irreversible MCMC method. 


	\subsubsection{Asymptotic variance}                                                     

    	\begin{figure*}[t]
    	\centering
		\includegraphics[width=.95\textwidth]{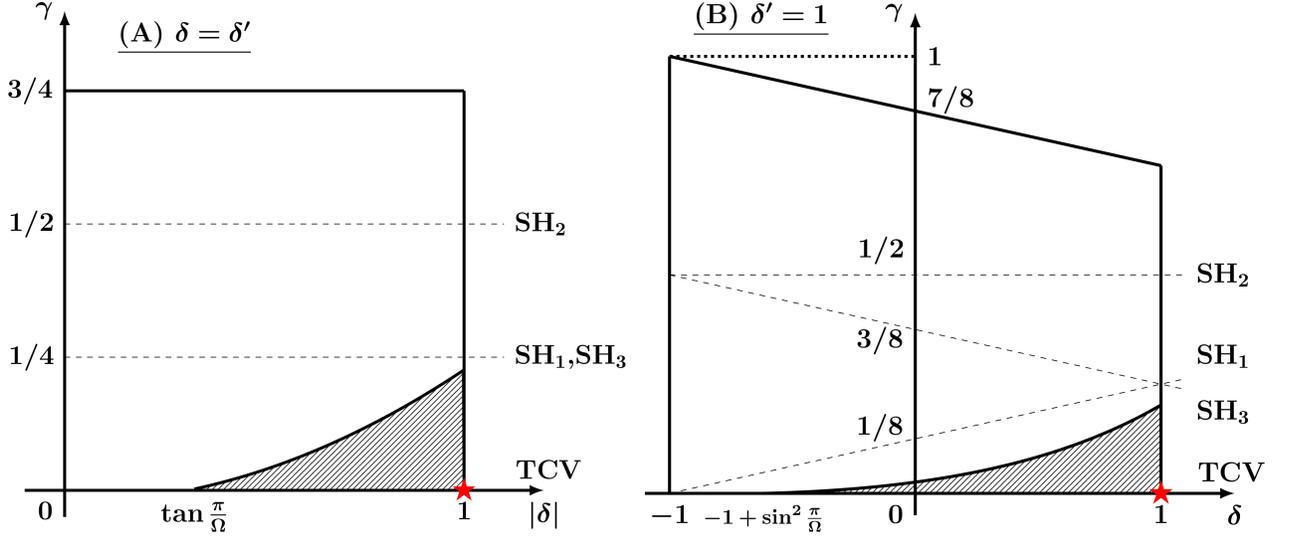}
		\caption{(Color online) The parameter region 
		$(\delta,\gamma)$ where the asymptotic variance is reduced by 
		imposing the skew detailed balance condition (shaded portion).
		The left and right panels represent cases (A) $\delta=\delta'$ and 
		(B) $\delta'=1$, respectively.
		The star symbol indicates the lowest value in the parameter region.}
		\label{fig:asymptotic-variance}
		\end{figure*}

	From the general result discussed in Sec.~\ref{Sect:MCwithDBC}, 
	the ratio of the asymptotic variance and the variance 
	for any quantity is upper-bounded by
	$(1+\lambda_2)/(1-\lambda_2)$
	for the reversible Markov chain.
	We study the corresponding upper bound of the ratio in the case
	of an irreversible random walk with the SDBC.
	It is reasonable to consider 
	the asymptotic variance for the extended quantity defined as
	$\hat{f}':(i,\varepsilon)\mapsto f_i$ because we are interested 
	in the quantities that are independent of the 
	auxiliary variable $\varepsilon$. Then, we obtain 
	the asymptotic variance of $\hat{f}'$ as
		\begin{align}
		&\qquad
		v(\hat{f}',\tilde{\bm{\pi}},\tilde{\mathsf{T}})
		\notag \\
		&=
		\sum_{k=2}^{[\Omega/2]+1}\left[\sum_{\varepsilon=\pm}
		\frac1{1-(A_k^{\varepsilon})^2}
		\frac{1+\tilde{\lambda}_k^{\varepsilon}}{
		1-\tilde{\lambda}_k^{\varepsilon}}\right]
		\left[\sum_{\sigma=1}^{m_k}
		(\bm{f}\mathsf{B}\bm{v}_{k,\sigma}^{\top})
		(\bm{u}_{k,\sigma}\bm{f}^{\top})\right].
		\end{align}
	The ratio of the asymptotic variance to the variance is found
	to be maximized when $\hat{f}'$ 
	is chosen as $\hat{f}'=\hat{v}'_{2,\sigma}$: 
		\begin{align}
		\max_{\hat{f}'\not=\hat{0}}
		\frac{v(\hat{f}',\tilde{\bm{\pi}}, \tilde{\mathsf{T}})}{
		{\rm var_{\tilde{\bm{\pi}}}}[\hat{f}']}
		&=
		\frac{v(\hat{v}_{2,\sigma}',\tilde{\bm{\pi}},\tilde{\mathsf{T}})}{
		{\rm var}_{\tilde{\bm{\pi}}}[\hat{v}'_{2,\sigma}]}
		\notag \\
		&=
		\sum_{\varepsilon=\pm}\frac1{1-(A_2^{\varepsilon})^2}
		\frac{1+\tilde{\lambda}_2^{\varepsilon}}{
		1-\tilde{\lambda}_2^{\varepsilon}}.
		\label{eq:upper bound of asymptotic vaiance 
		for extended physical quantity}
		\end{align}

	By comparing the upper bound of the asymptotic variance, 
	we show that the worst evaluation of the asymptotic variance of
	$\tilde{\mathsf{T}}$ 
	is improved from that of $\mathsf{T}$ 
	when the parameters $(\delta,\gamma)$ 
	satisfy the following inequalities: 
		\begin{align}
		\gamma\le\frac14\left(\delta^2\cos^2\frac{\pi}{\Omega}
		-\sin^2\frac{\pi}{\Omega}\right)
		\end{align}
	for case (A) and
		\begin{align}
		\gamma\le\frac{1+\delta}8\left(\frac{1+\delta}{3-\delta}
		\cos^2\frac{\pi}{\Omega}-\sin^2\frac{\pi}{\Omega}\right)
		\end{align}
	for case (B), respectively.
	The parameter region is illustrated in
	Fig.~\ref{fig:asymptotic-variance}.\ This result\ indicates 
	that the upper bound of the 
	asymptotic variance can never be improved by changing 
	the value of $\delta$ for the SH$_1$,  
	SH$_2$, or SH$_3$ type of transition rate. However, 
	by choosing the TCV type of transition rate, 
	it is improved when $\delta$ is sufficiently large and the limit of
	$|\delta|=1$ provides the most efficient algorithm in the
	parameter region.       
	The most efficient point is different from that in the sense of
	the relaxation time. 

        
	\section{\label{Sect:Summary and Discussion} Summary and Discussion}
        
	In this paper, the irreversible Metropolis-Hastings algorithm 
	satisfying the SDBC is generalized so that it is applicable 
	to any system with discrete degrees of freedom. 
	In this algorithm, the violation of the DBC is characterized by 
	the function $\Delta_{ij}^{(\pm)}$, referred to as the skewness function.
	In addition, there are four different transition probabilities 
	for $\varepsilon$ flip $\Lambda_i^{(\pm)}$ referred to as 
	the SH$_1$, SH$_2$, SH$_3$, and TCV types. 
	This algorithm has already been applied to several 
	Ising spin systems and the relaxation dynamics of magnetization density 
	has been discussed~\cite{YS2013,KH2013}.
        
	To acquire further knowledge about this algorithm, 
	it is applied to the random walk in one dimension as a benchmark.
	According to the general procedure mentioned in Sec.~\ref{Sect:MCwithSDBC},
	the irreversible transition matrix $\tilde{\mathsf{T}}$, 
	which is characterized by the parameters $\delta$, $\delta'$, and $\gamma$, 
	is constructed.
	Then, all the eigenvalues and eigenvectors of $\tilde{\mathsf{T}}$ are
 	derived analytically by the explicit diagonalization, and 
	the parameter $(\delta,\gamma)$ dependence of 
	the relaxation rate and the asymptotic variance are discussed 
	in the two particular cases $\delta=\delta'$ and $\delta'=1$.
	As a result, it is found that 
	the relaxation rate and the asymptotic variance in the irreversible 
	MCMC method are improved by selecting appropriate values of 
	parameters $(\delta,\gamma)$,
	in comparison with those in the corresponding reversible one.
	In particular, the relaxation rate is qualitatively improved by 
	the appropriate choice of the parameter $\gamma$.
	Therefore, it is theoretically confirmed that 
	the violation of the DBC by imposing the SDBC 
	can improve the efficiency of MCMC methods.
	From the present theoretical analysis of the toy model,
	it should be noticed that the efficiency of the MCMC method 
	depends on how its efficiency is evaluated, and the violation of the DBC 
	does not always improve the efficiency.

	As discussed in this paper, the efficiency of our proposed
	algorithm depends on the transition probability $\Lambda_i^{(\pm)}$.
	The choice of the skewness function $\Delta_{ij}^{(\pm)}$
	also affects the efficiency in general. 
	Only the simplest skewness function is 
	discussed in Sec.~\ref{Sect:Performance Evaluation},
	but other types of the skewness function are available such as
		\begin{align}
		\Delta_{ij}^{(\pm)}=\pm\delta{\rm tanh}[\beta(f_j-f_i)],
		\end{align}
	where $|\delta|\le 1$ and $\beta>0$ are parameters and 
	$f_i$ is the realization of quantity $\hat{f}$. 
	One may expect that there exists an appropriate choice of 
	skewness function that effectively reduces 
	the relaxation time and the asymptotic variance, depending on
	the measured quantity $\hat{f}$.
	It would be interesting 
	to clarify which choice is the most efficient 
	for each problem to be studied and also for each quantity to be measured. 
      
	Although in this paper we have explicitly shown that our
	proposed algorithm efficiently works in the simple random walk,
	little is known about its effectiveness in general.
	Thus, it is important to investigate what happens if we apply this method to 
	other statistical-mechanical models, such as Potts models, spin glasses, 
	and so on, which is left for future study.


	\begin{acknowledgments}
	YS is supported by a Grant-in-Aid for the
	Japan Society for Promotion of Science (JSPS) Fellows 
	(Grant No. 26$\cdot$7868). 
	KH is supported by Grants-in-Aid for Scientific Research from MEXT, 
	Japan (Nos. 25610102, and 25120010).
	\end{acknowledgments}
        

	\appendix

	\section{\label{sect:A}
	Verification of the irreversible Metropolis-Hastings algorithm}

	In this appendix, we verify that 
	a sequence $(\tilde{X}^{(n)})_{n=0,1,2,...}$ generated 
	with the IMH algorithm in Sec.~\ref{subsect:IMH} is a Markov chain 
	characterized by the transition matrix
	$\tilde{\mathsf{T}}$. It is obvious that the sequence
	$(\tilde{X}^{(n)})_{n=0,1,2,...}$ is a homogeneous Markov chain.  
	Thus, we only have to examine whether a conditional probability 
	${\rm Prob}[\tilde{X}^{(n+1)}=(j,\varepsilon')|
	\tilde{X}^{(n)}=(i,\varepsilon)]$ is identical with the
	corresponding element of $\tilde{\mathsf{T}}$.
	For $i\in I$, $j\not=i$, $\varepsilon=\pm$, and $n=0,1,2,...$, 
	each correspondence is verified by the following calculations:
		\begin{align}
		{\rm Prob}[\tilde{X}^{(n+1)}=(j,-\varepsilon)|
		\tilde{X}^{(n)}=(i,\varepsilon)]=0,
		\end{align}
		\begin{align}
		&
		{\rm Prob}[\tilde{X}^{(n+1)}=(j,\varepsilon)|
		\tilde{X}^{(n)}=(i,\varepsilon)]
		\notag \\
		&~=
		q_{ij}\frac{1+\Delta_{ij}^{(\varepsilon)}}{2}w_{ij}
		=
		T_{ij}^{(\varepsilon)},
		\end{align}
		\begin{align}
		&
		{\rm Prob}[\tilde{X}^{(n+1)}=(i,-\varepsilon)|
		\tilde{X}^{(n)}=(i,\varepsilon)]
		\notag \\
		&~=
		\left[q_{ii}+\sum_{j\not=i}q_{ij}
		\left(1-\frac{1+\Delta_{ij}^{(\varepsilon)}}{2}w_{ij}\right)\right]
		\frac{\Lambda_i^{(\varepsilon)}}{
		1-\sum_{j\not=i}T_{ij}^{(\varepsilon)}}
		\notag \\
		&~=
		\left(\sum_{j=1}^{\Omega}q_{ij}-\sum_{j\not=i}
		q_{ij}\frac{1+\Delta_{ij}^{(\varepsilon)}}{2}w_{ij}\right)
		\frac{\Lambda_i^{(\varepsilon)}}{
		1-\sum_{j\not=i}T_{ij}^{(\varepsilon)}}
		\notag \\
		&~=
		\left(1-\sum_{j\not=i}T_{ij}^{(\varepsilon)}\right)
		\frac{\Lambda_i^{(\varepsilon)}}{
		1-\sum_{j\not=i}T_{ij}^{(\varepsilon)}}=
		\Lambda_i^{(\varepsilon)},
		\end{align}
	and
		\begin{align}
		&
		{\rm Prob}[\tilde{X}^{(n+1)}=(i,\varepsilon)|
		\tilde{X}^{(n)}=(i,\varepsilon)]
		\notag \\
		&~=
		\left[q_{ii}+\sum_{j\not=i}q_{ij}
		\left(1-\frac{1+\Delta_{ij}^{(\varepsilon)}}{2}w_{ij}\right)\right]
		\notag \\
		&\qquad\qquad
		\times\left(1-\frac{\Lambda_i^{(\varepsilon)}}{
		1-\sum_{j\not=i}T_{ij}^{(\varepsilon)}}\right)
		\notag \\
		&~=
		1-\sum_{j\not=i}T_{ij}^{(\varepsilon)}-\Lambda_i^{(\varepsilon)}
		=
		T_{ii}^{(\varepsilon)}.
		\end{align}

        
	\section{\label{sect:B}Spectral decomposition of $\mathsf{J}_{\Omega}(a,b)$}

	In this appendix, the matrix $\mathsf{J}_{\Omega}(a,b)
	\in\mathbb{R}^{\Omega\times \Omega}$, defined by
		\begin{align}
		\mathsf{J}_{\Omega}(a,b)\equiv\left(
			\begin{matrix}
			0 & a &   &   & b \\
			b & 0 & a &   &   \\
			& \ddots & \ddots & \ddots &   \\
			  &   & b & 0 & a \\
			a &   &   & b & 0 
			\end{matrix}
		\right),
		\end{align}
	is considered. 
	By specifying all the eigenvalues and eigenvectors, 
	the spectral decomposition of $\mathsf{J}_{\Omega}(a,b)$ is derived.


	The eigenvalues of $\mathsf{J}_{\Omega}(a,b)$ with $a=b$ are given by
		\begin{align}
		\mu_k(a,b)
		=&
		(a+b)\cos\theta_k+i(a-b)\sin\theta_k,
		\label{eq:eigenvalue-J1}
 		\end{align}
	where $\theta_k\equiv2\pi(k-1)/\Omega$ for $k=1,2,\ldots,\Omega$. 
	The imaginary unit is denoted by $i$.
	The left (right) eigenvectors $\bm{u}_k$ ($\bm{v}_k$) 
	associated with the eigenvalue $\mu_k(a,b)$ are obtained as
		\begin{align}
		\bm{u}_k=\left(\frac1{\Omega}e^{-in\theta_k}\right)_{n=1}^{\Omega},\quad
		\bm{v}_k=\left(e^{in\theta_k}\right)_{n=1}^{\Omega},
		\end{align}
	respectively.
	They satisfy the orthonormal relation and the complete relation described as
		\begin{align}
		\bm{u}_k\bm{v}_l^{\top}=\delta_{kl}
		\end{align}
	and
		\begin{align}
		\sum_{k=1}^{\Omega}\bm{v}_k^{\top}\bm{u}_k=\mathsf{I}_{\Omega},
		\end{align}
	respectively. Thus, the spectral decomposition of $\mathsf{J}_{\Omega}(a,b)$ 
	is derived as
		\begin{align}
		\mathsf{J}_{\Omega}(a,b)=\sum_{k=1}^{\Omega}
		\mu_k(a,b)\bm{v}_k^{\top}\bm{u}_k.
		\end{align}
                

	For $a=b$, all the eigenvalues of $\mathsf{J}_{\Omega}(a,a)$ are real and 
	the eigenvalues $\mu_k(a,a)$ and $\mu_{\Omega+2-k}(a,a)$ 
	in Eq.~(\ref{eq:eigenvalue-J1}) are degenerate
	for $2\le k\le[(\Omega+1)/2]$ 
	because $\theta_{\Omega+2-k}=2\pi-\theta_k$.
	Thus, the eigenvalue of $\mathsf{J}_{\Omega}(a,a)$ is obtained as
		\begin{align}
		\mu_k(a,a)=2a\cos\theta_k,
		\end{align}
	for $k=1,2,\ldots,[\Omega/2]+1,$ and
	the multiplicity of $\mu_k(a,a)$ is given as
		\begin{align}
		\left\{
			\begin{array}{lcl}
			m_1=1 & & \text{for}~k=1,\\
			m_k=2 & & \text{for}~2\le k\le[(\Omega+1)/2],\\
			m_{\Omega/2+1}=1 & & \text{for}~k=\Omega/2+1~\text{with even}~\Omega.
			\end{array}
		\right.
		\end{align}     
	In this case, all the eigenvectors can be chosen as real vectors 
	because all the eigenvalues are real numbers.
	Let $\sigma=1,\ldots,m_k$ be an index for multiplicity 
	and $\bm{1}\equiv(1,\ldots,1)\in\mathbb{R}^{\Omega}$.
	Then, the left eigenvector $\bm{u}_{k,\sigma}$ and the right eigenvector 
	$\bm{v}_{k,\sigma}$ associated with 
	$\mathsf{J}_{\Omega}(a,a)$'s eigenvalue $\mu_k(a,a)$ 
	are given as follows: 
		\begin{align}
		\bm{u}_{1,1}=
		\frac1\Omega\bm{1},
		\quad
		\bm{v}_{1,1}=
		\bm{1},
		\end{align}
		\begin{align}
 		\left\{
			\begin{array}{l}
			\bm{u}_{k,1}=
			\left(\frac{\sqrt{2}}{\Omega}\cos{n\theta_k}\right)_{n=1}^{\Omega},
			\\
			\bm{v}_{k,1}=
			\left(\sqrt{2}\cos{n\theta_k}\right)_{n=1}^{\Omega},
			\end{array}
		\right.
		\\
		\left\{
			\begin{array}{l}
			\bm{u}_{k,2}=
			\left(\frac{\sqrt{2}}{\Omega}\sin{n\theta_k}\right)_{n=1}^{\Omega},
			\\
			\bm{v}_{k,2}=
			\left(\sqrt{2}\sin{n\theta_k}\right)_{n=1}^{\Omega},
			\end{array}
		\right.
		\end{align}
	for $2\le k\le[(\Omega+1)/2]$, and
		\begin{align}
		\left\{
			\begin{array}{l}
			\bm{u}_{\Omega/2+1,1}=
			\left(\frac1{\Omega}(-1)^n\right)_{n=1}^{\Omega},
			\\
			\bm{v}_{\Omega/2+1,1}=\left((-1)^n\right)_{n=1}^{\Omega},
 			\end{array}
		\right.
		\end{align}
	for $k=\Omega/2+1$ with even $\Omega$.          
	These eigenvectors satisfy the orthonormal relation 
		\begin{align}
		\bm{u}_{k,\sigma}\bm{v}_{l,\rho}^{\top}=\delta_{kl}\delta_{\sigma\rho}
		\end{align}
	and the complete relation
		\begin{align}
		\sum_{k=1}^{[\Omega/2]+1}\sum_{\sigma=1}^{m_k}
		\bm{v}_{k,\sigma}^{\top}\bm{u}_{k,\sigma}=\mathsf{I}_{\Omega}.
		\end{align}
	Thus, the spectral decomposition of $\mathsf{J}_{\Omega}(a,a)$ is derived as
 	\begin{align}
	\mathsf{J}_{\Omega}(a,a)=\sum_{k=1}^{[\Omega/2]+1}
	\mu_k(a,a)\left(\sum_{\sigma=1}^{m_k}
	\bm{v}_{k,\sigma}^{\top}\bm{u}_{k,\sigma}\right).
	\end{align}
                                

	\section{\label{sect:C}Spectral decomposition of
	$\tilde{\mathsf{J}}_{2\Omega}(a,b;c)$}
        
	The matrix $\tilde{\mathsf{J}}_{2\Omega}(a,b;c)
	\in\mathbb{R}^{2\Omega\times 2\Omega}$, defined as
		\begin{align}
		\tilde{\mathsf{J}}_{2\Omega}(a,b;c)\equiv\left(
			\begin{matrix}
			\mathsf{J}_{\Omega}(a,b) & c\mathsf{I}_{\Omega} \\
			c\mathsf{I}_{\Omega} & \mathsf{J}_{\Omega}(b,a)
			\end{matrix}
		\right),
		\end{align}
	is considered. In this appendix, all the eigenvalues and 
	eigenvectors of $\tilde{\mathsf{J}}_{2\Omega}(a,b;c)$ are explicitly given.
	It should be noted that
	the matrix $\tilde{\mathsf{J}}_{2\Omega}(a,b;c)$ is not diagonalizable 
	for $c=\pm(a-b)\sin\theta_k$ with $k=2,3,\ldots,[(\Omega+1)/2]$.

                        
	If $c=0$, the matrix $\tilde{\mathsf{J}}_{2\Omega}(a,b;0)$
	is equivalent to a block diagonal matrix 
	$\mathsf{J}_{\Omega}(a,b) \oplus\mathsf{J}_{\Omega}(b,a)$.
	Thus, the eigenvalue of $\tilde{\mathsf{J}}_{2\Omega}(a,b;0)$ is given as
		\begin{align}
		\tilde{\mu}_k(a,b;0)=(a+b)\cos\theta_k+i(a-b)\sin\theta_k,
		\end{align}
	for $k=1,2,\ldots,\Omega$, and all the eigenvalues are doubly-degenerate.
	Let $\bm{0}\equiv(0,\ldots,0)\in\mathbb{R}^{\Omega}$, 
	$\bm{u}_k$ and $\bm{v}_k$ be the vectors defined in Appendix \ref{sect:B},
	and an asterisk denote the complex conjugate.
	Then, the left eigenvector $\tilde{\bm{u}}_{k,\sigma}$ and 
	the right eigenvector $\tilde{\bm{v}}_{k,\sigma}$ associated with 
	$\tilde{\mu}_k(a,b;0)$ are straightforwardly obtained as
		\begin{align}
		\tilde{\bm{u}}_{k,1}=(\bm{u}_{k},\bm{0}),\quad
		\tilde{\bm{v}}_{k,1}=(\bm{v}_{k},\bm{0})
		\end{align}
	and
		\begin{align}
		\tilde{\bm{u}}_{k,2}=(\bm{0},\bm{u}^*_{k}),\quad
		\tilde{\bm{v}}_{k,2}=(\bm{0},\bm{v}^*_{k}),
		\end{align}
	for $1\le k\le \Omega$, respectively.
	They satisfy the orthonormal and complete relation and thus 
	the spectral decomposition of $\tilde{\mathsf{J}}(a,b;0)$ is 
	obtained as
		\begin{align}
		\tilde{\mathsf{J}}_{2\Omega}(a,b;0)=
		\sum_{k=1}^{\Omega}\tilde{\mu}_k(a,b;0)\left(
		\tilde{\bm{v}}_{k,1}^{\top}\tilde{\bm{u}}_{k,1}
		+\tilde{\bm{v}}_{k,2}^{\top}\tilde{\bm{u}}_{k,2}
		\right).
		\end{align}
        
        
	When $c\not=0$ and $a=b$, 
	the eigenvalue of $\tilde{\mathsf{J}}_{2\Omega}(a,a;c)$ is 
	given as
		\begin{align}
 		\tilde{\mu}_{k}^{\pm}(a,a;c)=2a\cos\theta_k\pm c,
		\end{align}
	for $k=1,2,\ldots,[\Omega/2]+1$.
	The multiplicity of $\tilde{\mu}_k^{\pm}(a,a;c)$,       
	which depends only on the label $k$, is obtained as
		\begin{align}
		\left\{
			\begin{array}{lcl}
			m_1=1 & & \text{for}~k=1,\\
			m_k=2 & & \text{for}~2\le k\le[(\Omega+1)/2],\\
			m_{\Omega/2+1}=1 & & \text{for}~k=\Omega/2+1~\text{with even}~\Omega.
			\end{array}
		\right.
		\end{align}     
	Let $\bm{u}_{k,\sigma}$ and $\bm{v}_{k,\sigma}$ 
	be the vectors defined in Appendix \ref{sect:B}.        
	Then, the left eigenvector $\tilde{\bm{u}}_{k,\sigma}^{\pm}$ and 
	the right eigenvector $\tilde{\bm{v}}_{k,\sigma}^{\pm}$ 
	associated with $\tilde{\mu}_k^{\pm}(a,a;c)$ are given as follows:
		\begin{align}
		\tilde{\bm{u}}_{1,1}^{\pm}=\frac1{2\Omega}(\bm{1},\pm\bm{1}),\quad
		\tilde{\bm{v}}_{1,1}^{\pm}=(\bm{1},\pm\bm{1}),
		\end{align}
	and
		\begin{align}
		\left\{
			\begin{array}{l}
			\tilde{\bm{u}}_{k,\sigma}^{\pm}=
			\frac1{\sqrt{2}}(
			\bm{u}_{k,\sigma},\pm\bm{u}_{k,\sigma}),
			\\
			\tilde{\bm{v}}_{k,\sigma}^{\pm}=
			\frac1{\sqrt{2}}(
			\bm{v}_{k,\sigma}, \pm\bm{v}_{k,\sigma}),
			\end{array}
		\right.
		\end{align}
	for $2\le k\le[\Omega/2]+1$.    
	They satisfy the orthonormal relation described as
		\begin{align}
		\tilde{\bm{u}}_{k,\sigma}^{\varepsilon}
		\tilde{\bm{v}}_{l,\rho}^{\mu\top}=
		\delta_{kl}\delta_{\sigma\rho}\delta_{\varepsilon\mu}
		\end{align}
	and the complete relation 
		\begin{align}
		\sum_{k=1}^{[\Omega/2]+1}\sum_{\varepsilon=\pm}\sum_{\sigma=1}^{m_k}
		\tilde{\bm{v}}_{k,\sigma}^{\varepsilon\top}
		\tilde{\bm{u}}_{k,\sigma}^{\varepsilon}=\mathsf{I}_{2\Omega}.
		\label{eq:complete-tildeT2}
		\end{align}
	Thus, the spectral decomposition of $\tilde{\mathsf{J}}_{2\Omega}(a,a;c)$
	is derived as
		\begin{align}
		\tilde{\mathsf{J}}_{2\Omega}(a,b;c)=\sum_{k=1}^{[\Omega/2]+1}\left[
		\sum_{\varepsilon=\pm}\tilde{\mu}_k^{\varepsilon}(a,a;c)
		\left(\sum_{\sigma=1}^{m_k}
		\tilde{\bm{v}}_{k,\sigma}^{\varepsilon\top}
		\tilde{\bm{u}}_{k,\sigma}^{\varepsilon}\right)\right].
		\end{align}
                        
        
	If $c\not=0$, $a\not=b$, and 
	$c\not=\pm(a-b)\sin\theta_k$ for all $k=2,3,\ldots,[(\Omega+1)/2]$,
	the eigenvalue of 
	$\tilde{\mathsf{J}}_{2\Omega}(a,b;c)$ is obtained as
		\begin{align}
		\tilde{\mu}_k^{\pm}(a,b;c)=
		(a+b)\cos\theta_k\pm\sqrt{c^2-(a-b)^2\sin^2\theta_k},
		\label{eq:B1}
		\end{align}
	for $k=1,2,\ldots,[\Omega/2]+1$. 
	Note that $\tilde{\mu}_k^{\pm}(a,b;c)$ might be complex when 
	$|c|<|a-b|\sin\theta_k$.
	The multiplicity of $\tilde{\mu}_k^{\pm}(a,b;c)$, 
	depending only on the label $k$, is given as
		\begin{align}
		\left\{
			\begin{array}{lcl}
			m_1=1 & & \text{for}~k=1,\\
			m_k=2 & & \text{for}~2\le k\le[(\Omega+1)/2],\\
			m_{\Omega/2+1}=1 & & \text{for}~k=\Omega/2+1~\text{with even}~\Omega.
			\end{array}
		\right.
		\end{align}
	The left eigenvector $\tilde{\bm{u}}_{k,\sigma}^{\pm}$ and 
	the right eigenvector $\tilde{\bm{v}}_{k,\sigma}^{\pm}$ 
	associated with $\tilde{\mu}_k^{\pm}(a,b;c)$ are given as follows:
		\begin{align}
		\tilde{\bm{u}}_{1,1}^{\pm}=\frac1{2\Omega}(\bm{1},\pm\bm{1}),\quad
 		\tilde{\bm{v}}_{1,1}^{\pm}=(\bm{1},\pm\bm{1}),
		\end{align}
		\begin{align}
		\left\{
			\begin{array}{l}
			\tilde{\bm{u}}_{k,1}^{\pm}=
			\frac1{\sqrt{2}[1-(A_k^{\pm})^2]}(
			\bm{u}_{k,1}+A_k^{\pm}\bm{u}_{k,2},
			\bm{u}_{k,1}-A_k^{\pm}\bm{u}_{k,2}),
			\\
			\tilde{\bm{v}}_{k,1}^{\pm}=
			\frac1{\sqrt{2}}(
			\bm{v}_{k,1}-A_k^{\pm}\bm{v}_{k,2}, 
			\bm{v}_{k,1}+A_k^{\pm}\bm{v}_{k,2}),
			\end{array}
		\right.
		\\
		\left\{
			\begin{array}{l}
			\tilde{\bm{u}}_{k,2}^{\pm}=
			\frac1{\sqrt{2}[1-(A_k^{\pm})^2]}(
			\bm{u}_{k,2}-A_k^{\pm}\bm{u}_{k,1},
			\bm{u}_{k,2}+A_k^{\pm}\bm{u}_{k,1}),
			\\
			\tilde{\bm{v}}_{k,2}^{\pm}=
			\frac1{\sqrt{2}}(
			\bm{v}_{k,2}+A_k^{\pm}\bm{v}_{k,1}, 
			\bm{v}_{k,2}-A_k^{\pm}\bm{v}_{k,1}),
			\end{array}
		\right.
		\end{align}
	for $2\le k\le[(\Omega+1)/2]$, where
		\begin{align}
		A_k^{\pm}\equiv\frac{c\mp\sqrt{c^2-(a-b)^2\sin\theta_k}}{
		(a-b)\sin\theta_k}
		\end{align}
	and
		\begin{align}
		\left\{
			\begin{array}{l}
			\tilde{\bm{u}}_{\Omega/2+1,1}^{\pm}=\frac1{\sqrt{2}}(
			\bm{u}_{\Omega/2+1,1},\pm\bm{u}_{\Omega/2+1,1}),
			\\
			\tilde{\bm{v}}_{\Omega/2+1,1}^{\pm}=\frac1{\sqrt{2}}(
 			\bm{v}_{\Omega/2+1,1},\pm\bm{v}_{\Omega/2+1,1}),
			\end{array}
		\right.
		\end{align}
	for $k=\Omega/2+1$ with even $\Omega$.
	They satisfy the orthonormal and complete relation described as
		\begin{align}
		\tilde{\bm{u}}_{k,\sigma}^{\varepsilon}
		\tilde{\bm{v}}_{l,\rho}^{\mu\top}=
		\delta_{kl}\delta_{\sigma\rho}\delta_{\varepsilon\mu}
		\end{align}
	and
		\begin{align}
  		\sum_{k=1}^{[\Omega/2]+1}\sum_{\varepsilon=\pm}\sum_{\sigma=1}^{m_k}
 		\tilde{\bm{v}}_{k,\sigma}^{\varepsilon\top}
		\tilde{\bm{u}}_{k,\sigma}^{\varepsilon}=\mathsf{I}_{2\Omega},
		\end{align}
	respectively. Thus, the spectral decomposition of 
	$\tilde{\mathsf{J}}_{2\Omega}(a,b;c)$ is obtained as
		\begin{align}
		\tilde{\mathsf{J}}_{2\Omega}(a,b;c)=\sum_{k=1}^{[\Omega/2]+1}
		\left[\sum_{\varepsilon=\pm}
		\tilde{\mu}_k^{\varepsilon}(a,b;c)\left(\sum_{\sigma=1}^{m_k}
		\tilde{\bm{v}}_{k,\sigma}^{\varepsilon\top}
		\tilde{\bm{u}}_{k,\sigma}^{\varepsilon}
		\right)\right].
		\label{eq:B-sp}
		\end{align}

        
	Let us consider the case in which 
	$a\not=b$ and there exists an integer $q\in\{2,3,\ldots,[(\Omega+1)/2]\}$ 
	such that $c=\epsilon(a-b)\sin\theta_q\equiv c_q$ with $\epsilon=\pm 1$.     
	In this case, the eigenvalues $\tilde{\mu}_q^+(a,b;c_q)$ and 
	$\tilde{\mu}_q^-(a,b;c_q)$ obtained from Eq.~(\ref{eq:B1}) 
	are degenerate. 
	Therefore, the multiplicity of the eigenvalue 
	$\tilde{\mu}_q\equiv \tilde{\mu}_q^{\pm}(a,b;c_q)$ is 4.
	However, the dimension of the eigenspace corresponding to $\tilde{\mu}_q$
	is 2 and thus $\tilde{\mathsf{J}}_{2\Omega}(a,b;c_q)$ 
	is not diagonalizable.
	In this case, one can transform $\tilde{\mathsf{J}}_{2\Omega}(a,b;c_q)$
	into a Jordan normal form by considering the set of vectors defined as
		\begin{align}
		\left\{
			\begin{array}{l}
			\tilde{\bm{u}}_{q}^{\rm (i)}=B_{q}^{\rm (i)}(
			\bm{u}_{q,1}-\epsilon\bm{u}_{q,2},
			\bm{u}_{q,1}+\epsilon\bm{u}_{q,2}),\\
			\tilde{\bm{u}}_{q}^{\rm (ii)}=B_{q}^{\rm (ii)}(
			\bm{u}_{q,1}+\epsilon\bm{u}_{q,2},
			\bm{u}_{q,1}-\epsilon\bm{u}_{q,2}),\\
			\tilde{\bm{u}}_{q}^{\rm (iii)}=B_{q}^{\rm (iii)}(
			\bm{u}_{q,2}+\epsilon\bm{u}_{q,1},
			\bm{u}_{q,2}-\epsilon\bm{u}_{q,1}),\\   
			\tilde{\bm{u}}_{q}^{\rm (iv)}=B_{q}^{\rm (iv)}(
			\bm{u}_{q,2}-\epsilon\bm{u}_{q,1},
			\bm{u}_{q,2}+\epsilon\bm{u}_{q,1})\\                   
			\end{array}
		\right.
		\end{align}
	and
		\begin{align}
		\left\{
			\begin{array}{l}
			\tilde{\bm{v}}_{q}^{\rm (i)}=C_{q}^{\rm (i)}(
			\bm{v}_{q,1}-\epsilon\bm{v}_{q,2},
			\bm{v}_{q,1}+\epsilon\bm{v}_{q,2}),\\
			\tilde{\bm{v}}_{q}^{\rm (ii)}=C_{q}^{\rm (ii)}(
			\bm{v}_{q,1}+\epsilon\bm{v}_{q,2},
			\bm{v}_{q,1}-\epsilon\bm{v}_{q,2}),\\
			\tilde{\bm{v}}_{q}^{\rm (iii)}=C_{q}^{\rm (iii)}(
			\bm{v}_{q,2}+\epsilon\bm{v}_{q,1},
			\bm{v}_{q,2}-\epsilon\bm{v}_{q,1}),\\   
			\tilde{\bm{v}}_{q}^{\rm (iv)}=C_{q}^{\rm (iv)}(
			\bm{v}_{q,2}-\epsilon\bm{v}_{q,1},
			\bm{v}_{q,2}+\epsilon\bm{v}_{q,1}),\\                   
			\end{array}
		\right.
		\end{align}
	where the coefficients $B_q^{\rm (x)}$ and $C_q^{\rm (x)}$ 
	(x $=$ i, ii, iii, iv) satisfy
		\begin{align}
		B_{q}^{\rm (x)}C_{q}^{\rm (x)}
		=&
		\frac14
		\end{align}
	and
		\begin{align}
		B_{q}^{\rm (ii)}C_{q}^{\rm (i)}
		=&
		B_{q}^{\rm (iv)}C_{q}^{\rm (iii)}=\frac12c_q.
		\end{align}
	Note that they satisfy the orthonormal relation as
		\begin{align}
		\tilde{\bm{u}}_q^{(\rm x)}\tilde{\bm{v}}_q^{(\rm y)\top}=\delta_{\rm xy}.
		\label{eq:B2}
		\end{align}
	Let $\mathsf{U},\mathsf{V}\in\mathbb{R}^{4\times 2\Omega}$ be
		\begin{align}
		\mathsf{U}\equiv\left(
			\begin{matrix}
			\tilde{\bm{u}}_{q}^{\rm (i)}\\
			\tilde{\bm{u}}_{q}^{\rm (ii)}\\
			\tilde{\bm{u}}_{q}^{\rm (iii)}\\
			\tilde{\bm{u}}_{q}^{\rm (iv)}
			\end{matrix}
		\right),\quad
		\mathsf{V}\equiv\left(
			\begin{matrix}
			\tilde{\bm{v}}_{q}^{\rm (i)}\\
 			\tilde{\bm{v}}_{q}^{\rm (ii)}\\
			\tilde{\bm{v}}_{q}^{\rm (iii)}\\
			\tilde{\bm{v}}_{q}^{\rm (iv)}
			\end{matrix}
		\right).
		\end{align}
	Then, the orthonormal relation in Eq.~(\ref{eq:B2}) can be rewritten as
		\begin{align}
		\mathsf{U}\mathsf{V}^{\top}=\mathsf{I}_4.
		\end{align}
	Moreover, by replacing the term 
		\begin{align}
		\sum_{\varepsilon=\pm}
		\tilde{\mu}_{q}^{\varepsilon}(a,b;c)
		\left(\sum_{\sigma=1,2}
		\tilde{\bm{v}}_{{q},\sigma}^{\varepsilon\top}
		\tilde{\bm{u}}_{{q},\sigma}^{\varepsilon}\right)
		\end{align}
	in Eq.~(\ref{eq:B-sp}) with 
		\begin{align}
		&\qquad \mathsf{V}^{\top}\left(
			\begin{matrix}
			\tilde{\mu}_q & 1 & & \\
			0 & \tilde{\mu}_q & & \\
			& & \tilde{\mu}_q & 1 \\
			& & 0 & \tilde{\mu}_q
			\end{matrix}
		\right)\mathsf{U}
		\notag \\
		=&\tilde{\mu}_q\sum_{\rm x}\tilde{\bm{v}}_{q}^{\rm (x)\top}
		\tilde{\bm{u}}_{q}^{\rm (x)}+
		\tilde{\bm{v}}_{q}^{\rm (i)\top}
		\tilde{\bm{u}}_{q}^{\rm (ii)}+
		\tilde{\bm{v}}_{q}^{\rm (iii)\top}
		\tilde{\bm{u}}_{q}^{\rm (iv)},
		\end{align}
	the spectral decomposition in Eq.~(\ref{eq:B-sp}) holds 
	even in this case.


\end{document}